\documentclass[10pt,aps,prx,reprint,superscriptaddress,floatfix,longbibliography]{revtex4-2}

\usepackage{physics}
\usepackage{amsmath}
\usepackage{amssymb}
\usepackage{amsthm}
\usepackage[dvipsnames]{xcolor}

\usepackage{amsbsy}
\usepackage{amstext}
\usepackage{graphicx}
\usepackage{color}

\usepackage{mathtools}
\usepackage{subfigure}
\usepackage{multirow}

\usepackage{amsfonts}
\usepackage{dcolumn}
\usepackage{bbold,bm}
\usepackage[bookmarks=false,linkcolor=green,urlcolor=blue,colorlinks,citecolor=red]{hyperref}

\makeatletter
\newcommand{\HH}{\mathcal{H}}
\newcommand{\tphi}{\widetilde{\varphi}}
\newcommand{\A}{\alpha}
\newcommand{\s}{\sigma}
\newcommand{\w}{\omega}
\newcommand{\D}{\Delta}
\newcommand{\kv}{\vb{k}}

\def\kt{\tilde{k}}
\def\E0{\varepsilon_0}
\makeatother
\usepackage{babel}

\begin{document}

\title{Photogalvanic response in multi-Weyl semimetals}

\author{Arpit Raj}
\email{raj.a@northeastern.edu}
\affiliation{Department of Physics, Northeastern University, Boston, MA 02115, USA}

\author{Swati Chaudhary}
\affiliation{Department of Physics, The University of Texas at Austin, Austin, Texas 78712, USA}
\affiliation{Department of Physics, Northeastern University, Boston, MA 02115, USA}
\affiliation{Department of Physics, Massachusetts Institute of Technology, Cambridge, MA 02139, USA}

\author{Gregory A. Fiete}
\affiliation{Department of Physics, Northeastern University, Boston, MA 02115, USA}
\affiliation{Department of Physics, Massachusetts Institute of Technology, Cambridge, MA 02139, USA}


\begin{abstract}
We investigate the dependence of the photogalvanic response of a multi-Weyl semimetal on its topological charge, tilt, and chemical potential. We derive analytical expressions for the shift and injection conductivities for tilted charge-$n$ Weyl points $(n=1,2,3)$ using a low energy two-band effective Hamiltonian. For double-Weyl semimetals, we also compute the response from two-band and four-band tight-binding models with broken time-reversal symmetry to study the effect of band bending and the contributions from higher bands. We find a significant deviation in the responses obtained from the effective low-energy continuum model and more realistic four-band continuum and tight-binding models. We analyze several different limits of these models. We describe the nature of the deviations and provide estimates of their dependence on the frequency and other model parameters. Our analysis provides a simple explanation for the first-principle calculation based frequency dependence of the injection current in SrSi$_2$. Additionally, we find interesting parameter regimes where the frequency dependence of the non-linear optical response can be directly used to probe the type-I/type-II nature of the Weyl cone.  We obtain analytical results for the charge-4 Weyl semimetal by reducing the original problem involving a triple $k$-space integral to one with only a double integral. This simplification allows us to extract all relevant information about the nature of its second-order dc response and the precise condition for observing circular photogalvanic effect quantization. The semi-analytical approach presented here can also be extended to a systematic study of second harmonic generation and first-order optical conductivity in charge-4 Weyl semimetals.
\end{abstract}
\maketitle

%
\section{Introduction}
The quantum geometry (QG) of Bloch wavefunctions can significantly influence the electronic properties and response functions of a material~\cite{Xiao2010}. The quantum anomalous Hall effect in the absence of a magnetic field is the seminal example of such QG effects, originating in this case from the Chern number~\cite{Haldane1988}. More generally, the anomalous contribution from band topology can overcome limitations in non-topological systems on many physical properties like superfluid weight~\cite{Xie2020}, exciton stability~\cite{Hu2022}, transport coefficients~\cite{Xiao2006}, and optical responses~\cite{ma2021topology,orenstein2021topology}. The bulk photovoltaic effect (BPVE) is one such effect where quantum geometry contributions have been shown to be of immense importance~\cite{sipe2000,dai2023recent}. The BPVE is a second-order optical response where a DC current is produced in response to an AC electric field. It has been shown that in many non-centrosymmetric materials, the non-trivial structure of Bloch wavefunctions engenders a BPVE without creating any macroscopic electric field or carrier concentration gradient in the sample~\cite{young2011}. This allows one to overcome the Shockley-Queisser limit~\cite{shockley1961detailed} present in traditional p-n junctions.

Based on the mechanism of generation, the bulk photovoltaic effects can be divided into shift and injection currents~\cite{sipe2000}. The shift current results from the real-space shift in the electron wavepacket due to inter-band photoexcitation~\cite{Fregoso2017}, and the injection current is caused by change in electron velocity upon inter-band transition~\cite{sipe2000}. The properties of these responses are determined by the polarization of light, and presence of time-reversal and space-inversion symmetries~\cite{Ahn2020PRX}. The shift current response occurs for linearly-polarized light even when time-reversal symmetry is present. On the other hand, the injection current requires  circularly polarized light and is also known as the circular photogalvanic effect (CPGE). However, when time-reversal symmetry is broken, both the shift current and injection current can  occur for circularly and linearly polarized light, respectively~\cite{PhysRevResearch.2.033100}.

These mechanisms for a BPVE are intimately related to the quantum geometry of the electronic wavefunction~\cite{ahn2022riemannian,morimoto2016topological}, and thus are proving to be reliable tools to probe and utilize the band topology~\cite{orenstein2021topology,ma2021topology}. The bulk photovoltaic effects in Weyl semimetals have attracted enormous research interest as they provide a mechanism to generate photocurrents in the infrared and THz regime~\cite{Chan2017,nagaosa2020transport}. It was shown in the seminal work~\cite{Juan2017}, that the CPGE contribution from a  Weyl node would exhibit quantization proportional to the charge of the Weyl node. Following these theory works, CPGE was measured in many different Weyl semimetals including  TaAs~\cite{ma2017direct}, RhSi~\cite{rees2020helicity}, and TaIrTe~$_4$\cite{ma2019nonlinear} which showed interesting helicity-dependent behavior arising from the chirality of Weyl nodes. These experimental works also highlighted the importance of using more realistic models: the tilt and higher bands were shown to play an important role in determining the CPGE response~\cite{Sadhukan2021a}. 
 
In recent years, many different kinds of Weyl semimetals have been discovered~\cite{Armitage2018,burkov2018weyl,PhysRevLett.119.206401,chang2018topological}. In certain materials, it has been shown that the Weyl node  carrying a charge higher than $n=1$ can be stabilized by crystal symmetries~\cite{Fang2012}. These semimetals, also known as multi-Weyl semimetals (MWSMs) have been proposed in SrSi$_2$~\cite{huang2016new,singh2018tunable}, Cu$_2$Se, and RhAs$_3$~\cite{Zhu2018} which can host Weyl nodes with charge $n=2$. It was shown in Ref.~\cite{Zhang2018eng,Ghorashi2018} that such double-Weyl nodes can also be engineered in Luttinger semimetals like $\alpha$-Sn by applying strain and magnetic fields or via Floquet engineering. 
 
Materials which can host Weyl nodes with Berry monopole charge higher than two are not known but triple-WSM can be possibly obtained from cubic Dirac semimetals~\cite{liu2017predicted} by applying a magnetic field or by Floquet engineering~\cite{hubener2017creating}. Another interesting feature of these multi-Weyl semimetals is that the dispersion around Weyl node is no longer linear in all directions but becomes quadratic (cubic) for two directions for charge two (three). This leads to a strong anisotropy in the velocity matrix and also modifies the density of states which has been known to affect the transport coefficients~\cite{Chen2016MWSM,Fu2022,menon2020anomalous,sinha2019transport,Kulikov2020,Park2017,dantas2018magnetotransport,nag2020magneto,Huang2017,fu2022thermoelectric,gorbar2017anomalous,roy2022non,Zeng2023,PhysRevB.105.214307,bouhlal2022tunnelingwiley,bouhlal2021tunnelingscience} and linear optical responses~\cite{ahn2017optical,Sun20178,mukherjee2018,yadav2022magneto,ghosh2022theoretical,Das2022} of multi-Weyl semimetals. These unusual properties of multi-Weyl semimetals are also believed to significantly influence the second-order optical responses, such as the BPVE and second-harmonic generation. A deeper understanding of how different properties of these MWSMs affect the shift current and injection current can possibly lead to a mechanism to probe the topological charge of Weyl semimetals.  
 
Most theoretical works on the BPVE employ effective two-band low-energy Hamiltonians. These models have proven quite useful for general predictions like the quantization of the injection current conductivity, but the experimental signatures are often complicated by the discrepancy between effective low-energy models and real electronic band structure where the band curvatures and higher energy bands start to play an important role. As a result, the predictions of the continuum model usually agree only in a small energy window. This necessitates the need to analyze the role of different model parameters and understand the frequency behavior of Weyl semimetals in different regimes away from this small energy window. 

In our work, we first provide a complete analytical solution to the two-band charge-$n$ low energy Hamiltonian along with an analysis of its important features, including CPGE quantization. These analytical expressions elucidate the role of tilt and non-linear dispersion on different components of the shift and injection current conductivities in multi-Weyl semimetals. We also numerically evaluate the response in tight-binding models and observe a significant deviation in some components of second-order conductivity which highlight the importance of band curvature.  

For multi-Weyl semimetals, the validity of two-band models becomes further restricted. Double Weyl nodes are obtained when two charge-1 Weyl nodes  are pinned to a high-symmetry point and two of the four bands are gapped out by some symmetry allowed perturbations. As a result, even if the effective two-band picture is valid for each charge-1 Weyl node in a given energy range, it might not be valid for the double-Weyl node if the perturbation is not strong enough to push the other two bands out of that energy window. This type of scenario occurs in the charge-2 WSM SrSi$_2$~\cite{huang2016new} where the two charge-1 Weyl nodes are gapped out by a spin-orbit coupling resulting in a very small gap between the bands hosting a double-Weyl node and the higher energy bands. 

Inspired by the band structure of SrSi$_2$, we also consider a four-band continuum model and find a significant deviation from the two-band continuum model. We find that the CPGE quantization is destroyed and instead a very different behavior is observed at small frequencies. In the particular case of the four-band model, we  find two opposite limits in the parameter space with good and poor agreement. We notice that the agreement is better when the perturbation induced gap is large. Our analysis provides a simple-explanation for the results from first-principle calculations in Ref.~\cite{Sadhukhan2021} where quantization is observed only above a certain cutoff frequency. We attribute this discrepancy to the contribution from higher bands. 

Finally, we also investigate the charge-4 case by using a two-band effective low energy model and a tight-binding model. We derive semi-analytical expressions for different components of the shift and injection current conductivities. Most importantly, we obtain the analytical limits for the frequency window where CPGE quantization can be observed.

Our paper is organized as follows. In Sec.II, we provide a brief introduction to the shift and injection current conductivities along with the symmetry requirements to observe their effects. In Sec.III, we derive expressions for different components of these second-order conductivity tensors by considering an effective two-band low-energy Hamiltonian for a Weyl node with arbitrary charge $n$. We also include a finite tilt in the $z$-direction in our analysis and systematically study how tilt affects these different components at different chemical potentials and frequencies. In Sec.IV, we focus on double-Weyl semimetals and consider two different models. First, we compare different conductivities for a two-band tight-binding model and an effective low-energy Hamiltonian. Next, we consider a four-band model inspired by the SrSi$_2$ band structure around its double Weyl node and study the second-order conductivities in different limits. In Sec.V, we derive the joint density of states (JDOS), and the shift and injection current conductivity expressions for a charge-4 model. In Sec.VI, we discuss the implications of our results.

%
\section{Photogalvanic response}
In materials lacking inversion symmetry, the photogalvanic effect (PGE) refers to the generation of directed photocurrent as a second-order response to an external time-varying electromagnetic field. For light of frequency $\w$ (and wavelength much larger than the sample size so the electric field has uniform amplitude), the second-order dc response is given by,
\begin{align}
 j_{dc}^a &= \s^{abc}(\omega) E_b(\w)E_c(-\w),
\end{align}
where the second-order conductivity $\sigma^{abc}(\omega)$ can be divided into a shift current conductivity,  $\s_{\text{shift}}^{abc}$ and an injection current conductivity, $\s_{\text{inj}}^{abc}$. These two quantities are given by,
\begin{align}
\begin{split}
 \s_{\text{shift}}^{abc} &= \frac{-i\pi e^3}{\hbar^2} \int_{\kv}\sum_{n>m} f_{nm} \Big( r_{nm}^b r_{mn;a}^{c} - r_{mn}^c r_{nm;a}^b \Big) \\
 & \hspace{4.5cm} \times \delta (\w_{nm}-\w),
 \end{split} \label{eq:shift_main}\\
 \s_{\text{inj}}^{abc} &= \tau\frac{2\pi e^3}{\hbar^2} \int_{\kv} \sum_{n>m} f_{nm}\D_{nm}^a r_{nm}^br_{mn}^c  \delta (\w_{nm}-\w), \label{eq:inj_main}
\end{align}
where, $n,m$ label the energy bands, $\int_{\kv} = \int \dd[3]{k}/(2\pi)^3$, $\w_{nm}=\w_n-\w_m$ is the energy difference between bands $n$ and $m$, $f_{nm}=f_n-f_m$ where $f$ is the Fermi-Dirac distribution function, $\D_{nm}^a = v^a_{nn}-v^a_{mm}$ with $v^a_{nn}$ being the velocity matrix elements, and $\tau$ is the relaxation time. The interband Berry connection is given by $r_{nm}^b = \bra{n}i\pdv{k_b}\ket{m} \text{ for } n\neq m$ and zero otherwise, with its generalized derivative defined as $r_{nm;a}^b = \pdv{r_{nm}^b}{k_a}-i(\xi_{nn}^a-\xi_{mm}^a)r_{nm}^b$, where $\xi_{nn}^a=\bra{n}i\pdv{k_a}\ket{n}$ is the intraband Berry connection. 

Numerical calculation of these quantities by direct evaluation of wavefunction derivatives can be difficult as it would require fixing a smooth gauge for the wavefunctions at each point. However, it is possible to circumvent this problem completely by making use of $r_{nm}^b=-iv_{nm}^b/\w_{nm}=-i\bra{n}\pdv{k_b}\HH\ket{m}/\w_{nm}$, and the sum rule~\cite{sipe2000,cook2017design,Ahn2020PRX},
\begin{align}
\begin{split}
 r_{nm;a}^b &= \frac{i}{\w_{nm}}\Bigg[\frac{\D_{nm}^bv_{nm}^a + \D_{nm}^av_{nm}^b}{\w_{nm}} - w_{nm}^{ba} \\
 &\qquad + \sum_{l\neq n,m} \bigg(\frac{v_{nl}^bv_{lm}^a}{\w_{lm}}-\frac{v_{nl}^av_{lm}^b}{\w_{nl}}\bigg)\Bigg], \quad n\neq m
\end{split} 
\end{align}
where, $w_{nm}^{ba}=\bra{n}\pdv{k_b}\pdv{k_a}\HH\ket{m}$. The condition on the summation $\sum_{l\neq n,m}$ is understood as $\w_l\neq \w_n,\w_m$~\cite{Ahn2020PRX}. 

The consequences of time-reversal symmetry can be seen directly by analyzing the integrand in Eq.~\eqref{eq:shift_main} and Eq.~\eqref{eq:inj_main} under a time-reversal operation. Time-reversal symmetry enforces the real part of the integrand to be odd in $\vec{k}$ space, and hence makes $\sigma^{abc}_\text{shift}$ real and $\sigma^{abc}_\text{inj}$ imaginary~\cite{zhang2018}. In other words, when time-reversal symmetry is preserved, the shift current conductivity is non-zero only for linearly-polarized light and the injection current requires circularly polarized light. However, no such restrictions are present once the time-reversal symmetry is broken.

%
\section{Results for the charge-$n$ low-energy Weyl Hamiltonian}
We begin with a low-energy effective Hamiltonian for a two-band charge-$n$ Weyl point
\begin{align}
 \HH_n &= \mqty( u_zk_z + u_tk_z - \mu    & \E0(\kt_x-i\zeta\kt_y)^n \\
              \E0(\kt_x+i\zeta\kt_y)^n   & - u_zk_z + u_tk_z - \mu), \label{eq:Hn}
\end{align}
where $\zeta=\pm1$, $u_z$ and $u_t$ are, respectively, the effective velocity and tilt along $\vu{z}$.  Here, $\kt_{x,y}=k_{x,y}/k_0$, and $\mu$ is the chemical potential. The values $k_0, \E0$ are material-dependent parameters with units of momentum and energy, respectively. We will assume $\E0>0$ and set $k_0=1$. The chirality of this Weyl point is $\chi=\text{sgn}(u_z\zeta)$. The energy eigenvalues are given by,
\begin{align}
 E_{n,\pm} &= u_tk_z - \mu\pm\E0\sqrt{(\kt_x^2+\kt_y^2)^n +  u_z^2k_z^2/\E0^2}.
\end{align}
It should be noted that although all our derivations will hold for $n$ being any positive integer, it makes physical sense to only take $n=1,2,3$ due to symmetry restrictions in actual lattice systems~\cite{zhang2020twofold,yu2022encyclopedia}. Two-band charge-4 Weyl points are allowed but have different low energy Hamiltonian~\cite{zhang2020twofold,cui2021charge4} and are discussed in a later section. 

In order to use Eq.~\eqref{eq:shift_main}, and Eq.~\eqref{eq:inj_main} to find the shift and injection conductivity tensors, we note that the delta and Fermi-Dirac distribution functions restrict the domain of integration. In our calculations, we assume temperature, $T=0$ K which simplifies the Fermi-Dirac distribution to $f(E)=1-\Theta(E)$ where $\Theta$ is the Heaviside  function. The delta function forces the integration to be performed over the surface $2\E0\sqrt{(\kt_x^2+\kt_y^2)^n + u_z^2k_z^2/\E0^2}-\w=0$, while the theta function further selects out a portion of this surface. By making suitable substitutions, this surface can be transformed into a sphere which makes it easier to perform the integral analytically (see Appendix \ref{appendix:cn}) for arbitrary charge $n$. 

\begin{figure}[h!]
\includegraphics[scale=0.55]{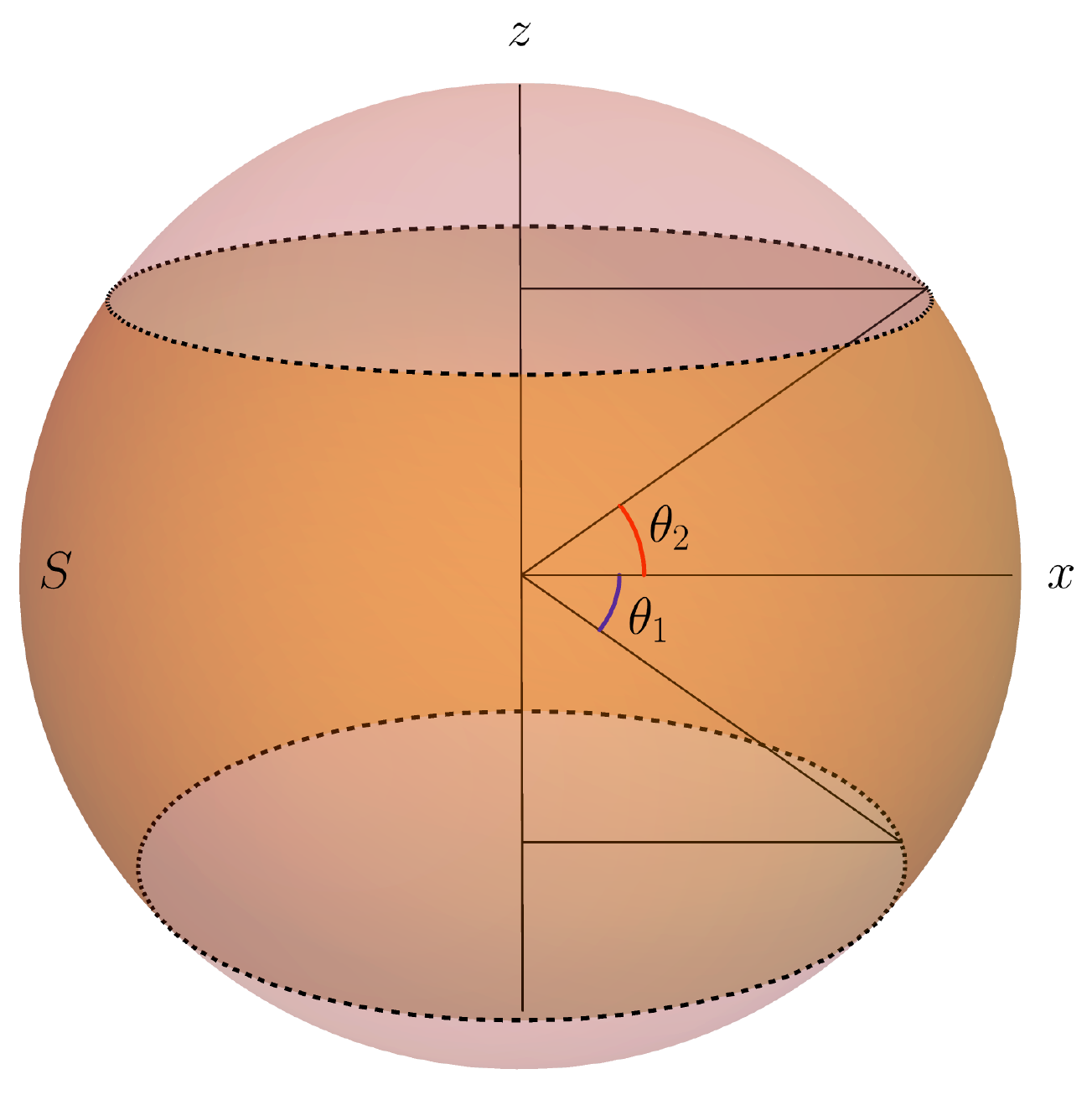}
\caption{The surface defined by $\delta (\w_{21}-\w)$ in the transformed coordinates (see Appendix~\ref{appendix:cn}). The factor $f_{21}$ restricts the integral in Eq.\eqref{eq:shift_main}, Eq.\eqref{eq:inj_main}, Eq.\eqref{eq:jdos} to the Pauli-unblocked region S (shown in brown).}
\label{fig:cn_sphere}
\end{figure}

After accounting for the finite tilt of the Weyl cone, the Pauli blocking condition restricts the integration region on this sphere  to region $S$ as shown in Fig.~\ref{fig:cn_sphere} with  $\theta_1$ and $\theta_2$ given by:
\begin{align}
 \theta_p &= \begin{dcases}
              -\pi/2, \text{ if } \varphi_p<-1 \\
              \arcsin(\varphi_p), \text{ if } -1\leq\varphi_p\leq 1 \\
              +\pi/2, \text{ if } 1<\varphi_p
             \end{dcases},
\label{thetaeq}
\end{align}
for $p=1,2$
where $\varphi_p=\frac{1}{W}\left(\text{sgn}\left(\frac{u_t}{u_z}\right)\frac{2\mu}{\w}+(-1)^p\right)$, and $W=|u_t/u_z|$ is an important quantity  which determines if the WSM is type-I ($W<1$) or type-II ($W>1$). The behavior of $\theta_1, \theta_2$ is mainly determined by the amount of tilt $(W)$ and doping $(\mu)$, and is crucial to understanding the basic features of the response. 

For zero doping, $\theta_2=-\theta_1=\pi/2$ for type-I and $\theta_2=-\theta_1=\arcsin(1/W)$ for type-II WSM. It should also be noted that the angles lose dependence on chirality in this case. It is important to note that these results contain an implicit $\w$ dependence. In the transformed coordinates, where the integration surface is a sphere, these angles are measured from the $x$-axis in the $xz$-plane and determine which part of that surface is not Pauli-blocked (region $S$ in Fig.~\ref{fig:cn_sphere}). 

First, we evaluate the join density of states using the expression
\begin{align}
 \text{JDOS}(\w) &= \int_{\kv} \sum_{n>m} f_{mn}  \delta (\w_{nm}-\w), \label{eq:jdos}
\end{align}
where the factor $f_{nm}$ accounts for Pauli-blocking effects. In the absence of the tilt, we obtain the expected $\omega^{2/n}$ dependence for a charge-$n$ Weyl node. However, at finite tilt and finite chemical potential, this $\w^{2/n}$ dependence is modulated by the angular factor of $\int_{\theta_1}^{\theta_2} \cos^{2/n-1}{\theta} \dd{\theta}$.

\renewcommand{\arraystretch}{1}
\begin{table*}[t]
\caption{Results for the low-energy charge-$n$ Hamiltonian in Eq.\eqref{eq:Hn} with $\E0,\w>0$. Note that $\chi=\text{sgn}(u_z\zeta)$.}
\begin{ruledtabular}
\begin{tabular}{|m{0.32\textwidth}|m{0.6\textwidth}|}
\hline
 $\mathrm{JDOS}$ & $\frac{k_0^2}{8\pi^2| u_z|}\frac{1}{n}\left(\frac{\w}{2\E0}\right)^{2/n} \int_{\theta_1}^{\theta_2} \cos^{2/n-1}{\theta} \dd{\theta}$ \\
\hline
 Shift conductivity   &   \\
\hline
  $\s^{xzx} = -\s^{xxz} = \s^{yzy} = -\s^{yyz}$ & $n\frac{i\mathrm{sgn}( u_z)e^3k_0^2}{32\pi\hbar^2} \big(\sin^2{\theta_2} - \sin^2{\theta_1}\big)\frac{1}{\w}$ \\
\hline
  $\s^{xyz} = \s^{xzy} = -\s^{yzx} = -\s^{yxz}$ & $n\frac{\mathrm{sgn}( u_z \zeta)e^3k_0^2}{32\pi\hbar^2} \big(\sin{\theta_2}\cos^2{\theta_2} - \sin{\theta_1}\cos^2{\theta_1}\big)\frac{1}{\w}$ \\
\hline
 Injection conductivity   &   \\
\hline
 $\s^{zxy} = -\s^{zyx}$ & $n\frac{i\tau\mathrm{sgn}( u_z\zeta)e^3k_0^2}{24\pi\hbar^2} \big(\sin^3{\theta_1} - \sin^3{\theta_2}\big)$ \\
\hline
 $\s^{yzy} = \s^{yyz} = \s^{xzx} = \s^{xxz}$ & $n\frac{\tau\mathrm{sgn}( u_z)e^3k_0^2}{64\pi\hbar^2} \big(\cos^4{\theta_1} - \cos^4{\theta_2}\big)$\\
\hline
 $\s^{xyz} = -\s^{xzy} = \s^{yzx} = -\s^{yxz}$ & $n\frac{i\tau\mathrm{sgn}( u_z\zeta)e^3k_0^2}{48\pi\hbar^2} \big(3\sin{\theta_1}-\sin^3{\theta_1} - 3\sin{\theta_2}+\sin^3{\theta_2}\big)$ \\
\hline
 $\s^{zxx} = \s^{zyy}$ & $n\frac{\tau\mathrm{sgn}( u_z)e^3k_0^2}{256\pi\hbar^2} \big(-6\cos(2\theta_1)+\cos^2(2\theta_1) + 6\cos(2\theta_2)-\cos^2(2\theta_2)\big)$ \\
\hline
 $\s^{zzz}$ & $ \frac{\tau u_z^2\mathrm{sgn}( u_z)e^3k_0^2}{2^{2+2/n}\E0^{2/n}(1+n)\pi\hbar^2 \w^{2-2/n}} \big(\cos^{2+2/n}{\theta_2} - \cos^{2+2/n}{\theta_1}\big)$ \\
\hline
\end{tabular}
\end{ruledtabular}
\label{tab:cn}
\end{table*}
\renewcommand{\arraystretch}{1}

Next, we calculate different components of shift and injection current tensors. The resulting expressions are given in Table~\ref{tab:cn}. We notice that all  conductivity tensors are directly proportional to the charge of the Weyl point, except for $\s^{zzz}_{\text{inj}}$. It should be noted that analytical results for the $n=1$ and untilted $n=2$ cases have been given in Refs.~\cite{cook2017design, juan2020difference, yang2017divergent, Ahn2020PRX} and Ref.~\cite{mandal2020effect}, respectively. Here, we have extended the analytical results to arbitrary chiral charge-$n$ with finite tilt.

Let's first analyze the shift current conductivity results.  As shown in Table~\ref{tab:cn}, there are two kinds of non-zero components: (i) purely imaginary which is responsible for a second-order dc photocurrent from circular polarization, and (ii) purely real which leads to a photogalvanic effect from linearly polarized light. For the shift current, the circular polarization components always vanish at zero doping since $\theta_1=-\theta_2$ for $\mu=0$ from Eq.~\eqref{thetaeq}. Similarly, when time-reversal symmetry (TRS) is preserved, the circular polarization current from a time-reversed pair of nodes would also vanish as $u_z\rightarrow -u_z$ under time-reversal. 

On the other hand, the linear polarization component $\sigma^{xyz}_\text{shift}$ shows a very interesting behavior and can even provide estimates of tilt and chemical potential. We note that among all the non-zero conductivity tensors, $\s^{xyz}_{\text{shift}}$ alone changes sign with frequency and can be  used to estimate $\mu$. For type-I and type-II with $W<2$, this sign change occurs at $\omega=2|\mu|$ which can be understood from Eq.~\eqref{thetaeq} which indicates that while one of the angles is zero the other becomes $\pm \pi/2$ leading to $\s^{xyz}_{\text{shift}}=0$. The latter stays at $\pm\pi/2$ for small variation in $\w$, while the former changes sign going through $\w=2|\mu|$, causing $\s^{xyz}_{\text{shift}}$ to do the same (as it has a $\sin{\theta}\cos^2{\theta}$ dependence).

\begin{figure*}[t!]
\includegraphics[scale=0.58]{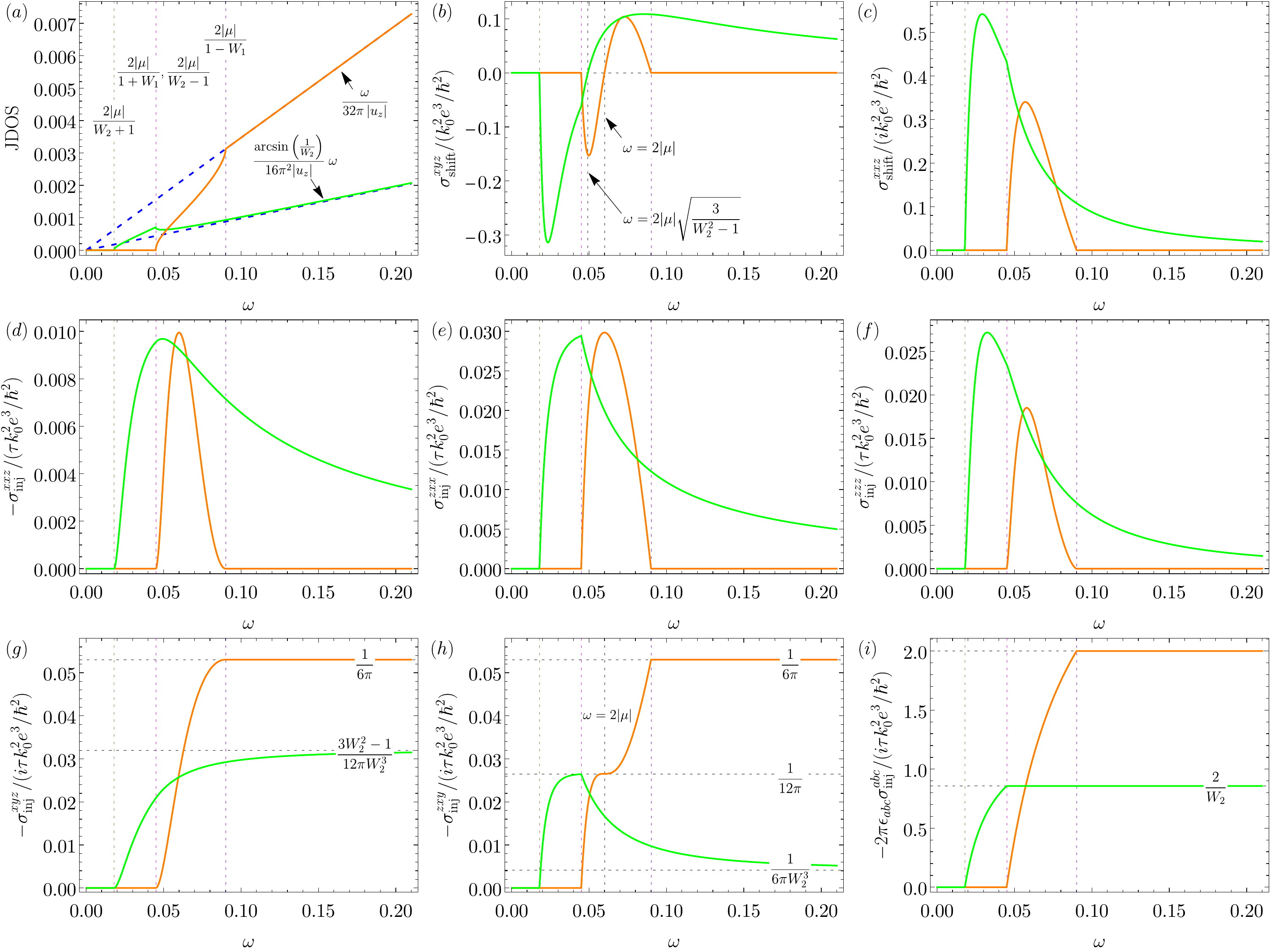}
\caption{Plots showing (a) JDOS, (b)-(c) shift conductivity,  (d)-(h) injection conductivity, (i) CPGE quantization for a single charge-2 Weyl point (obtained using the expressions in Table~\ref{tab:cn}). The orange and green curves correspond to type-I ($W_1=0.334$) and type-II ($W_2=2.334$) case, respectively. We have chosen $W_2-W_1=2$ just to keep the plots neat. We have taken $u_z=0.287,\E0=1,\mu=-0.03$. Note that except for the JDOS, remaining plots will show similar behavior for charges 1 and 3.}
\label{fig:c2-2bkp}
\end{figure*}

The $W\geq 2$ case is not so straightforward but after some work we find that the sign change occurs at $2|\mu|\sqrt{\tfrac{3}{W^2-1}}$ (see Appendix \ref{appendix:signchange} for details). Note that $\tfrac{2|\mu|}{1+W}< 2|\mu|\sqrt{\tfrac{3}{W^2-1}} \leq 2|\mu|$ with the equality holding at $W=2$, as one would expect. Interestingly, for $\mu=0$, both components $\s^{xyz}_\text{shift}, \s^{xxz}_\text{shift}$ show a $1/\w$ divergence for a type-II WSM. Additionally, for type-I, all shift current conductivities are non-zero (shown in Fig.~\ref{fig:c2-2bkp}(b,c)) only in a finite frequency window determined by the tilt parameter $W$ and doping. 

Our results show that the tilt parameter plays an important role for all shift current components. When the tilt vanishes, all the shift current conductivity components also vanish. This can be easily understood from the behavior of $\theta_p$ from Eq.\eqref{thetaeq} in the limit $W\rightarrow 0$. For $\w<2|\mu|$, $\theta_1=\theta_2=\pm\pi/2$ which simply means that the entire $\w_{21}=\w$ surface is Pauli-blocked. However, when $\w>2|\mu|$, the entire surface becomes Pauli-unblocked (as captured by $\theta_2=-\theta_1=\pi/2$) which again leads to a vanishing shift conductivity.

Now, we turn our attention to injection current conductivity components--some of which are known to exhibit quantization proportional to the Berry charge of the Weyl node. Here again, there are two kind of components: (i) purely imaginary which leads to CPGE, and (ii) purely real which leads to a photogalvanic effect from linearly polarized light. When time-reversal symmetry is preserved, the contribution from time-reversed Weyl node pairs is such that the real components vanish and only CPGE survives, as expected. Also, all the real components of the injection current conductivity would disappear at zero doping and also at zero tilt. As a result in order to get an injection current for linearly polarized light not only time-reversal must be broken but doping and tilt should be finite as well. 

For finite doping, the conductivities become non-zero after $2|\mu|/(1+W)$ for both type-I and type-II WSMs (note that the $1/\w$ divergence gets cut off in case of type-II). For type-I, $\s^{xyz}_\text{inj}, \s^{yzx}_\text{inj}, \s^{zxy}_\text{inj}$ reach their quantized value of $-n\,\text{sgn}(u_z\zeta)/12\pi$ after $2|\mu|/(1-W)$ where as other components become zero beyond this point. For the latter, the response window is proportional to $|\mu|$. For type-II case, $\s^{xyz}_\text{inj}, \s^{yzx}_\text{inj}, \s^{zxy}_\text{inj}$ approach their respective quantized values $-n\,\text{sgn}(u_z\zeta)\frac{3W^2-1}{12\pi W^3}$, $-n\,\text{sgn}(u_z\zeta)\frac{3W^2-1}{12\pi W^3}$, $\frac{-n\,\text{sgn}(u_z\zeta)}{6\pi W^3}$ asymptotically while the remaining components asymptotically approach zero.

The quantization condition for CPGE for the injection current conductivity can be easily obtained as the trace of the CPGE tensor,
\begin{align}
\frac{2\pi}{i\tau e^3k_0^2/\hbar^2} \epsilon_{abc}\s^{abc}_{\text{inj}} &= -n\,\mathrm{sgn}( u_z\zeta ) \left(\frac{\sin{\theta_2} - \sin{\theta_1}}{2}\right),
\end {align}
which gives the perfect quantized value equal  $-n\,\mathrm{sgn}(u_z\zeta)$ only when $\theta_2=-\theta_1=\pi/2$. The contribution of the  factor $\tfrac{1}{2}(\sin{\theta_2} - \sin{\theta_1})$ is easy to understand when interpreted as the fraction of the solid angle available for integration,
\begin{align}
\begin{split}
    \frac{1}{4\pi}\int_{S}\dd{\Omega} &= \frac{1}{4\pi}\Big(4\pi-\big(2\pi(1-\sin{\theta_2}) \\
    &\hspace{2.5cm} + 2\pi(1+\sin{\theta_1})\big)\Big), \\
\end{split}
\end{align}
which leads to a reduced value of quantized response when either $|\theta_1|,|\theta_2|<\pi/2$. For $\mu=0$, type-I WSM gives perfect quantization where as in type-II WSM, the quantization value is reduced by a factor of $1/W$. When $\mu\neq 0$, type-I WSMs show perfect quantization above a certain frequency cutoff, i.e for $\w\geq 2|\mu|/(1-W)$ where as type-II WSMs show a reduced quantization for $\w\geq 2|\mu|/(W-1)$, as shown in Fig.~\ref{fig:c2-2bkp}(i). Note that in the case of type-II, while individual terms in the CPGE trace only approach their respective quantized values asymptotically, the trace itself is fully quantized for $\w>2|\mu|/(W-1)$. This feature is captured in Fig.~\ref{fig:c2-2bkp}(g,h,i).

When TRS is broken, injection current can also be generated by linearly polarized light, and the non-zero components for this case depend on the tilt direction. For tilt along the $z$-axis, non-zero linear photogalvanic effect (LPGE) injection current conductivities include $\s^{yzy}, \s^{yyz}, \s^{xzx}, \s^{xxz}, \s^{zxx}, \s^{zyy},$ and $\s^{zzz}$. The last one is the only component among all shift and injection current conductivities which can allow for a current in the direction of linear polarization if it coincides with the direction of the tilt. Thus, a measurement of $\s^{aaa}_\text{inj}$  can provide a simple way to determine the direction of the tilt in charge 2 and 3 which have linearly dispersing bands in only one direction. 

%
\section{Charge-2 Weyl semimetals}
For concreteness, we numerically calculate the conductivity tensors for the following two-band tight-binding model for a charge-2 WSM with broken inversion, time-reversal and mirror symmetries,
\begin{eqnarray}
 \HH^{2b} &=& t\big(2(\cos(k_y)-\cos(k_x))\s_x + 2\sin(k_x)\sin(k_y)\s_y \nonumber \\
   &&\quad + (M-\cos(k_x)-\cos(k_y)-\cos(k_z))\s_z \nonumber \\
   &&\quad +g\sin{k_z}\s_0\big) - \mu\s_0,
\label{eq:2btbc2}
\end{eqnarray}
which has nodes at $(k_x,k_y,k_z)=(0,0,\pm\acos(M-2))$ for $1\leq M \leq 3$. The low-energy Hamiltonian near the nodes is given by,
\begin{align}
\begin{split}
\HH^{2b}_\pm &= t\big((k_x^2-k_y^2)\s_x + 2k_xk_y\s_y + u_z k_z\s_z \\
             &\quad + \left(u_t k_z - (\mu/t-g u_z)\right)\s_0\big),
\end{split} \label{eq:2bkpc2}
\end{align}
where, $u_z = \pm\sqrt{(3-M)(M-1)}$, and $u_t = g(M-2)$. Chirality of this node is given by $\chi=\text{sgn}(u_z)$. The bands disperse as $t(u_t\pm u_z)k_z$ when $k_x,k_y=0$ and $\pm t(k_x^2+k_y^2)$ when $k_z=0$. Based on the possible range of band inversion strength and $k$-space node separation given in Ref.~\cite{singh2018tunable}, we take $t=1\text{eV},\, g=0.1,\, M=2.958$ and $\mu=-0.0287t$. The chosen parameters give $W=0.33$ and put the lower energy node near zero energy. 

\begin{figure*}[t!]
\includegraphics[scale=0.58]{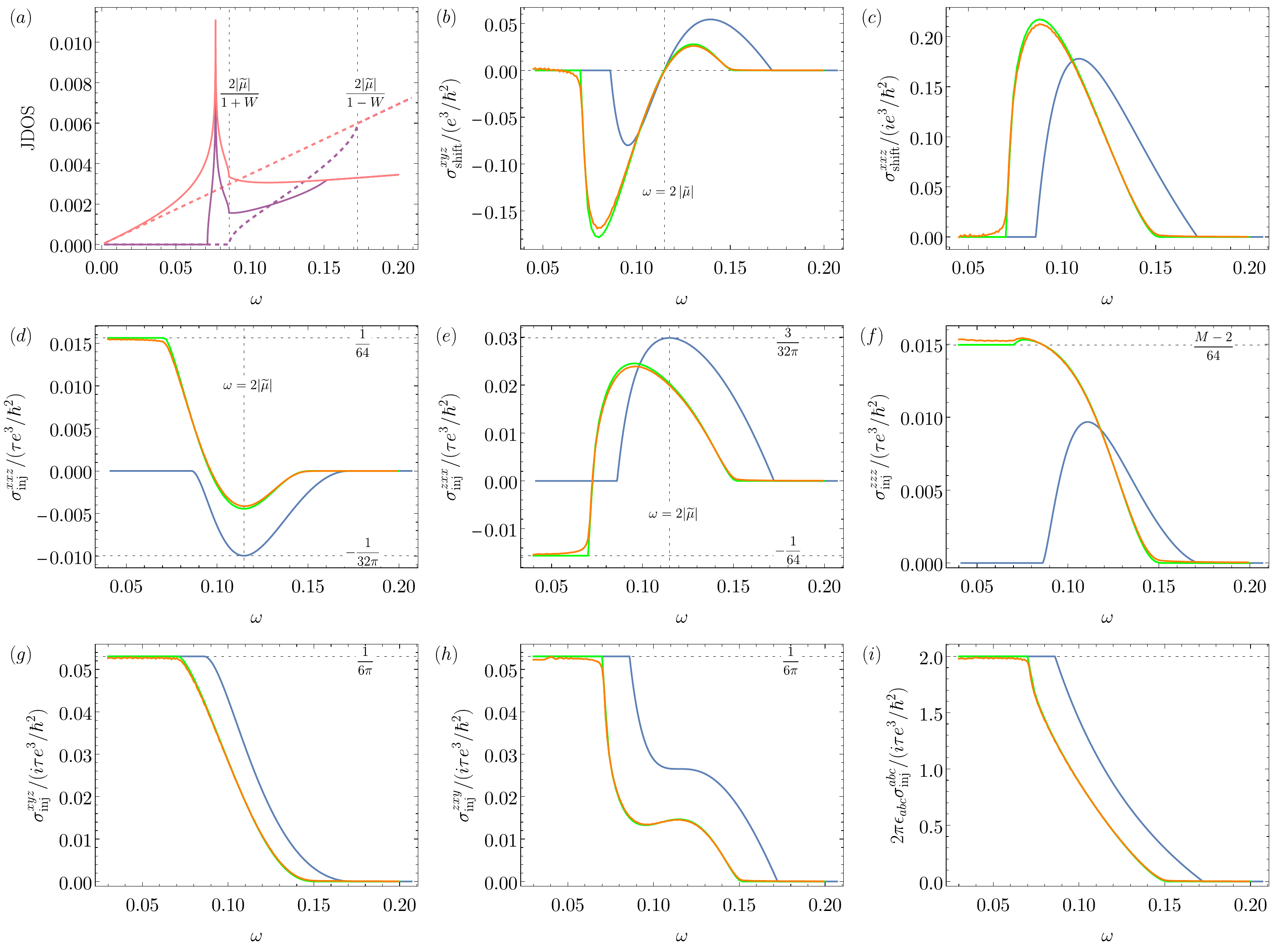}
\caption{Injection and shift conductivities for type-I charge-2 WSM. The orange curve represents the response for the two-band tight binding model Eq.\eqref{eq:2btbc2} obtained numerically. The blue curve is sum of contributions from the two nodes based on analytical results for the low energy model Eq.\eqref{eq:2bkpc2} while the green curve is obtained by including higher order terms $(\tfrac{1}{2}(k_x^2+k_y^2) + (\tfrac{1}{2}M-1) k_z^2)\s_z - \tfrac{1}{2}g u_z k_z^2 \s_0$ in Eq.\eqref{eq:2bkpc2} (see Appendix \ref{appendix:hoc} for details). The energy separation between the nodes is $|2g u_z|$ and $\widetilde{\mu}=\mu-gu_z$. We have taken $t=1, M=2.958, g=0.1, \mu=-0.0287$. (a) JDOS for each node for the low energy model with (solid) and without (dashed) higher order correction terms. Pink and purple correspond to the nodes at $-\mu-g u_z$ and $-\mu+g u_z$, respectively.}
\label{fig:c2-2btb}
\end{figure*}

The second-order dc conductivity results obtained for Eq.\eqref{eq:2btbc2} are shown in orange in Fig.~\ref{fig:c2-2btb}. To compare these results against those obtained simply by treating each node separately based on Eq.\eqref{eq:2bkpc2}, we have included the blue curve which represents the sum total of contributions from individual nodes using expressions from Table~\ref{tab:cn}. This is reasonable if the contribution from at least one of the nodes is constant over the energy range under consideration, as is the case here.

Looking at Fig.~\ref{fig:c2-2btb}, it is clear that higher order terms present in the tight binding model lead to significant deviations from the low-energy predictions of Table~\ref{tab:cn}. Surprisingly, we find that the injection conductivities $\s^{xxz},\s^{zxx}$ and $\s^{zzz}$ (d-f), develop a plateau up to an energy of about $0.07t$. Additionally, a shift in the response energy window to the left by about $0.02t$ is seen for all the conductivities (b-h) and the CPGE quantization (i). A shift in the quantization window has also been seen for $\mathcal{T}$-symmetric charge-2 Weyl system~\cite{Sadhukhan2021} and is believed to arise from higher order terms.

We probe the origin of these deviations by explicitly adding higher order terms to Eq.~\eqref{eq:2bkpc2} (see Appendix \ref{appendix:hoc} for details). Specifically, we find that including the second order terms $(\tfrac{1}{2}(k_x^2+k_y^2) + (\tfrac{1}{2}M-1) k_z^2)\s_z - \tfrac{1}{2}g u_z k_z^2 \s_0$ in Eq.\eqref{eq:2bkpc2}, not only matches the energy shift, but also captures all other features of the tight-binding results. Most importantly, we find the plateaus to come from the node situated close to zero energy (in our case, it is the node with negative $u_z$) and their heights to be $\frac{-\text{sgn}(u_z)}{64}$, $\frac{\text{sgn}(u_z)}{64}$ and $\frac{\text{sgn}(u_z)(2-M)}{64}$, respectively. 

We should point out that the $\s^{xxz}, \s^{zxx}$ plateaus can be obtained by just including the $\tfrac{1}{2}(k_x^2+k_y^2)\s_z$ term where as the one for $\s^{zzz}$ can be explained with the $(\tfrac{1}{2}M-1) k_z^2)\s_z$ term alone. We believe that the plateaus should be present in any charge-2 WSM where these higher order terms show up.
Note that despite the energy shift, $\s^{xyz}_\text{shift}$ crosses zero at $2|\mu-gu_z|$ as seen from Fig~\ref{fig:c2-2btb}(b). This may not hold for arbitrarily chosen $\mu$. In our case, we have carefully put one node close to zero energy which makes the other node almost entirely dictate the behavior of the response. 

Lastly, we also find perfect CPGE quantization up to an energy of about $0.07t$ as shown in Fig~\ref{fig:c2-2btb}(i). Behavior of the JDOS for each node with (solid) and without (dashed) higher order corrections is shown in Fig.~\ref{fig:c2-2btb}(a). With the higher order terms, the JDOS for the node at energy $-\mu+gu_z$ becomes non-zero after about $0.071t$ and explains why the quantization ceases earlier than the predicted value $0.086t$.

%
Beyond two-band models, the CPGE quantization is no longer guaranteed to hold. This has been explored in Ref.~\cite{Juan2017} for a charge-1 WSM. In order to better understand the contributions coming from higher bands and the extent to which they destroy quantization in charge-2 WSM, we study the following four-band tight-binding model (taking inspiration from Ref.~\cite{huang2016new}) with broken time-reversal symmetry,
\begin{align}
\begin{split}
 \HH^{4b} &= t\big(\sin(k_x)\tau_x + \sin(k_y)\tau_y \\
          &\quad + (M-\cos(k_x)-\cos(k_y)-\cos(k_z))\tau_z \\
          &\quad + \Delta\left(\tau_x\s_x+\tau_y\s_y\right) + g \sin(k_z) \tau_z\s_z\big) - \mu,
\end{split} \label{eq:4btbc2}
\end{align}
where $\tau$ and $\s$ are Pauli matrices acting on the orbital and spin space, respectively.
%
\begin{figure*}[t!]
\includegraphics[scale=0.58]{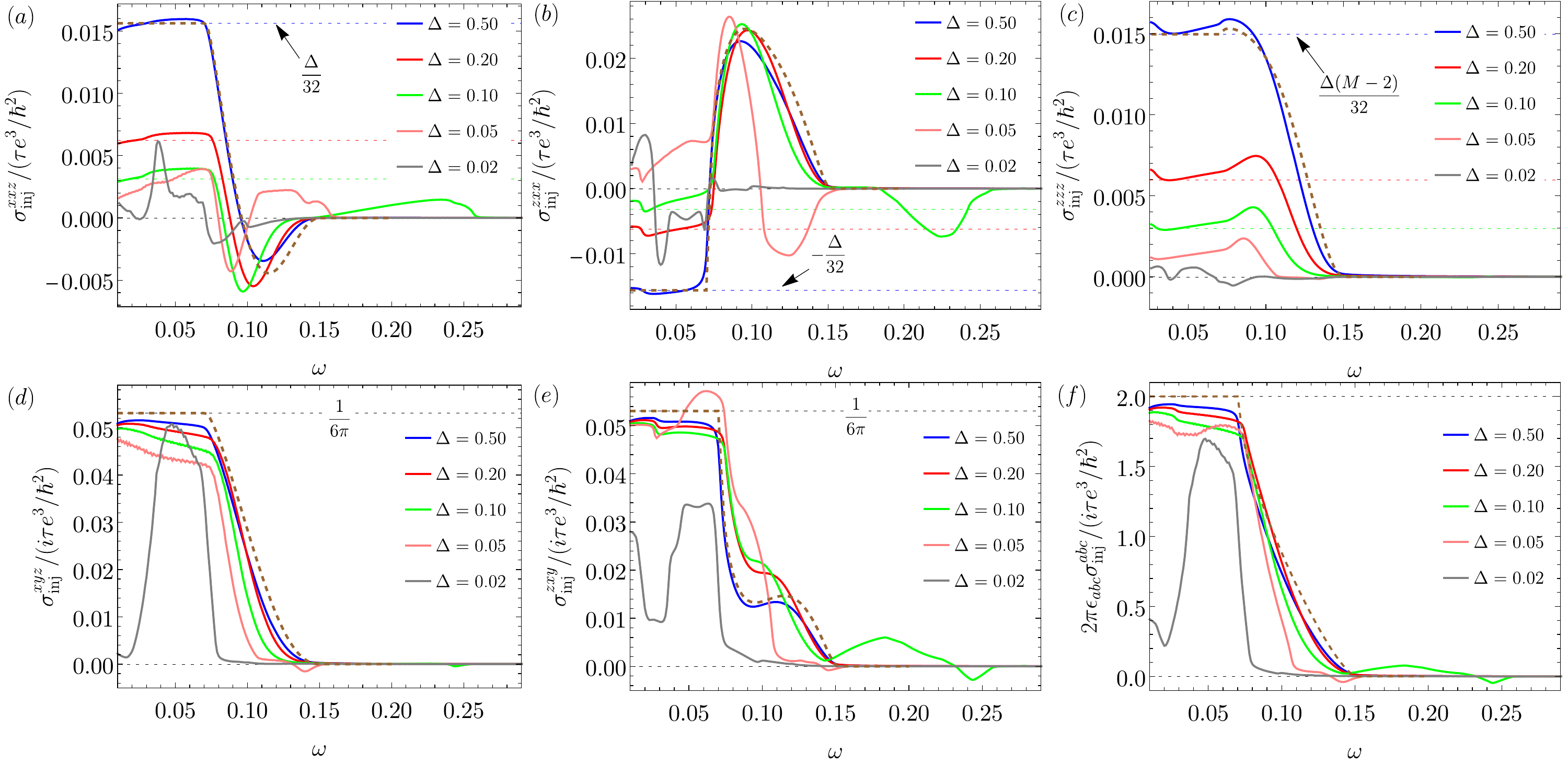}
\caption{(a)-(e) Injection conductivities and (f) trace of the CPGE tensor for charge-2 WSM obtained using the four-band model Eq.\eqref{eq:4btbc2}. The dashed brown curve shows the corresponding result for the two-band model (green curve from Fig.~\ref{fig:c2-2btb}) which is close to the $\Delta=0.5$ curve, as expected. We see significant deviations from perfect CPGE quantization for $\Delta\lesssim |gu_z|$. The $\s^{xxz}_\text{inj}, \s^{zxx}_\text{inj}, \s^{zzz}_\text{inj}$ plateaus continue to show up for $\Delta\gg |gu_z|$ with heights of about $\Delta/32$, $-\Delta/32$ and $\Delta(M-2)/32$, respectively.}
\label{fig:c2-4btb}
\end{figure*}

The $k_z$ dependent $\tau_z\s_z$ term produces tilt while $\Delta(\tau_x\s_x+\tau_y\s_y)$ gives rise to the quadratic band dispersion along $k_x,k_y$. The low-energy Hamiltonian near nodes at $(0,0,\pm\acos(M-2))$ is,
\begin{align}
\begin{split}
 \HH_{\pm}^{4b} &= t\big(k_x\tau_x + k_y\tau_y + u_zk_z\tau_z + \Delta\left(\tau_x\sigma_x+\tau_y\sigma_y\right) \\
 &\quad + (u_tk_z + g u_z)\tau_z\s_z\big) - \mu,
\end{split} \label{eq:4bkpc2}
\end{align}
where $u_z = \pm\sqrt{(3-M)(M-1)}$, and $u_t = g(M-2)$. 

We begin with $|\Delta|$ large compared to $|gu_z|$ and gradually decrease it to below $|gu_z|$. The two bands which touch, disperse as $t(u_t+u_z)k_z$, $t(u_t-u_z)k_z$ when $k_x,k_y=0$ and $\frac{|t|(k_x^2+k_y^2)}{2|\Delta+gu_z|}$, $\frac{-|t|(k_x^2+k_y^2)}{2|\Delta-gu_z|}$ when $k_z=0$. We use the same $g,M,t$ values from before. We note that unlike the two-band case Eq.\eqref{eq:2bkpc2}, the quadratic dispersion now has a dependence on $\Delta$. For $\Delta=0.5$ (recall $gu_z=0.0287$), the dispersion becomes almost the same for the two cases and provides a good starting point for comparison. Also, since the gap between the highest occupied and the lowest unoccupied bands is $\sim|\Delta|$, the effect of higher bands should be more prominent for smaller values of $\Delta$. 

Results obtained for Eq.\eqref{eq:4btbc2} are shown in Fig.~\ref{fig:c2-4btb}. We find large deviation from perfect quantization for small gaps as seen in Fig.~\ref{fig:c2-4btb}(f). However, for $\Delta\gg |gu_z|$ we do see almost perfect quantization. Also, the plateaus seen earlier in $\s^{xxz}_\text{inj}, \s^{zxx}_\text{inj}, \s^{zzz}_\text{inj}$ continue to show up when $\Delta$ is at least a few times larger than $|gu_z|$ as shown in (a-c). Their heights become dependent on $\Delta$ and are empirically found to be about $\Delta/32$, $-\Delta/32$, and $\Delta(M-2)/32$, respectively.

%
\section{Charge-4 Weyl semimetals}
%

\renewcommand{\arraystretch}{1.75}
\begin{table*}[t!]
    \centering
    \caption{Results for JDOS and injection conductivity for the Hamiltonian in Eq.\eqref{eq:2bkpc4} with $c_1,\w >0$ (when $c_1<0$, replace $c_1\rightarrow|c_1|$, $\mu\rightarrow -\mu$ and use the $c_1>0$ results). Integrals given here are to be evaluated over the region satisfying $x>0, z>0, \left(4 c_2^2 \left(x^2-x z+z^2\right)-c_1^2\right) \left(-c_3^2 x^2 \omega  z-8 c_2^2 c_1 (x-2 z)^2+8 c_1^3\right)>0$. Here, we have defined $ F_{\frac{2\mu}{\w}}(x,z)=\Theta\left(-x-z+1+\frac{2\mu}{\w}\right) \Theta\left(x+z+1-\frac{2\mu}{\w}\right) \Theta\left(1+\frac{2\mu}{\w}\right) $.}
    \begin{ruledtabular}
    \begin{tabular}{|c|c|}
    \hline
        JDOS & $\int_0^{1+\frac{2\mu}{\w}}\dd{z}\int_0^{1+\frac{2\mu}{\w}}\dd{x}\frac{c_1 \sqrt{\omega } F_{\frac{2\mu}{\w}}(x,z) }{4 \pi ^3 \sqrt{z} \sqrt{\left(4 c_2^2 \left(x^2-x z+z^2\right)-c_1^2\right) \left(-c_3^2 x^2 \omega  z-8 c_2^2 c_1 (x-2 z)^2+8 c_1^3\right)}}$\\
        \hline
        Injection conductivity &  \\
        \hline
        $\s^{xyz} = -\s^{xzy}$ & \multirow{2}{*}{$\int_0^{1+\frac{2\mu}{\w}}\dd{z}\int_0^{1+\frac{2\mu}{\w}}\dd{x} \frac{-i\tau e^3 \sqrt{3} c_2^2 c_3 \sqrt{\omega} \left(4 c_2^2 c_3^2 x \omega  z^2 \left(x^2-z^2\right)+c_3^2 c_1^2 x \omega  z^2-96 c_2^4 c_1 x z^2 (x-2 z)-16 c_2^2 c_1^3 (x+z)^2+4 c_1^5\right) F_{\frac{2\mu}{\w}}(x,z)}{2 \pi^2 \hbar^2 c_1^3 \sqrt{z} \left(24 c_1 c_2^2-c_3^2 \omega  z\right) \sqrt{\left(4 c_2^2 \left(x^2-x z+z^2\right)-c_1^2\right) \left(-c_3^2 x^2 \omega  z-8 c_2^2 c_1 (x-2 z)^2+8 c_1^3\right)}}$} \\
        $\s^{yzx} = -\s^{yxz}$ & \\
        \hline
        $\s^{zxy} = -\s^{zyx}$ & $\int_0^{1+\frac{2\mu}{\w}}\dd{z}\int_0^{1+\frac{2\mu}{\w}}\dd{x} \frac{-i\tau e^3 \sqrt{3} c_2^2 c_3 \sqrt{\omega } x z^2 \left(4 c_3^2 c_2^2 \omega  \left(z^2-x^2\right)+96 c_1 c_2^4 (x-2 z)+c_1^2 c_3^2 \omega \right) F_{\frac{2\mu}{\w}}(x,z)}{\pi^2 \hbar^2 c_1^3 \sqrt{z} \left(24 c_1 c_2^2-c_3^2 \omega  z\right) \sqrt{\left(4 c_2^2 \left(x^2-x z+z^2\right)-c_1^2\right) \left(-c_3^2 x^2 \omega  z-8 c_2^2 c_1 (x-2 z)^2+8 c_1^3\right)}}$\\
        \hline
        $\frac{2\pi}{i\tau e^3/\hbar^2} \epsilon_{abc}\s^{abc}$ & $\int_0^{1+\frac{2\mu}{\w}}\dd{z}\int_0^{1+\frac{2\mu}{\w}}\dd{x} \frac{8 \sqrt{3} c_2^2 c_3 \sqrt{\omega } \left(-2 c_1^3+8 c_1 c_2^2 (x+z)^2-c_3^2 x \omega  z^2\right) F_{\frac{2\mu}{\w}}(x,z)}{\pi  c_1 \sqrt{z} \left(24 c_1 c_2^2-c_3^2 \omega  z\right) \sqrt{\left(4 c_2^2 \left(x^2-x z+z^2\right)-c_1^2\right) \left(-c_3^2 x^2 \omega z - 8 c_2^2 c_1(x-2 z)^2 + 8 c_1^3\right)}}$ \\
    \hline
    \end{tabular}
    \end{ruledtabular}
    \label{tab:c4}
\end{table*}
\renewcommand{\arraystretch}{1}
%
Having looked at the charge-2 case in some detail, we move on to investigate the behaviour of the injection conductivity and JDOS for charge-4 WSMs. The existence of CPGE quantization in such systems has been discussed in earlier studies~\cite{cui2021charge4}. In our study, we want to develop a full understanding of how model parameters and doping affect these responses. In order to do that, we take the following two-band Hamiltonian based on Ref. \cite{cui2021charge4}, 
\begin{align}
\begin{split}
 \HH_4 &= -2c_1(\cos(k_x)+\cos(k_y)+\cos(k_z))\s_0 \\
            &\quad + 2c_2\big(\sqrt{3}(\cos(k_y)-\cos(k_x))\s_x \\
            &\quad - (\cos(k_x)+\cos(k_y)-2\cos(k_z))\s_z\big) \\
            &\quad + c_3\sin(k_x)\sin(k_y)\sin(k_z)\s_y - \widetilde{\mu}\s_0,
\end{split} \label{eq:2btbc4}
\end{align}
which has nodes of opposite chirality at $(0,0,0)$ and $(\pi,\pi,\pi)$. The low-energy Hamiltonian near $\Gamma$-point is given by,
\begin{align}
\begin{split}
 \HH_4^{\Gamma} &= c_1\left(k_x^2+k_y^2+k_z^2\right)\s_0 + c_2\Big(\sqrt{3}\left(k_x^2-k_y^2\right)\s_x \\
       &\quad + \left(k_x^2+k_y^2-2k_z^2\right)\s_z\Big) + c_3 k_xk_yk_z\s_y - \mu\s_0,
\end{split} \label{eq:2bkpc4}
\end{align}
where $\mu=\widetilde{\mu}+6c_1$. The chirality of this Weyl point is given by $\chi=\text{sgn}(c_3)$. We derive all our results using Eq.~\eqref{eq:2bkpc4} with $c_1>0$ (the opposite case is an easy generalization, which we discuss later). Its eigenvalues are given by
\begin{align}
\begin{split}
 E^{\Gamma}_{4,\pm} &= c_1\left( k_x^2+k_y^2+k_z^2 \right) \pm 2|c_2|\bigg( k_x^4+k_y^4+k_z^4 \\
           &\quad -k_x^2k_y^2-k_y^2k_z^2-k_z^2k_x^2 + \left(\frac{c_3}{2c_2}\right)^2 k_x^2k_y^2k_z^2 \bigg)^\frac{1}{2} - \mu.
\end{split}
\end{align}

The presence of the sixth order term $k_x^2k_y^2k_z^2$ above does not allow us to fully evaluate Eq.\eqref{eq:shift_main}, Eq.\eqref{eq:inj_main}, and Eq.\eqref{eq:jdos} analytically. However, it is possible to integrate out $k_y$ (one can pick any one out of the three $k$ coordinates) and get rid of the delta function in exchange for a new constraint (see Appendix \ref{appendix:c4} for details). 

The biggest advantage of going from a triple to a double integral is that the new constraint now defines a closed area compared to a closed surface before, which makes it much easier to analyze. The expressions for JDOS and injection conductivities (non-zero components) thus obtained are given in Table \ref{tab:c4}. Note that the shift conductivities are zero. Although these integrals appear complicated, they are easy to evaluate numerically. A key result of our analysis is the precise location of the energy window and condition under which trace of the CPGE tensor is quantized for different amounts of doping.

\begin{figure*}[t!]
\includegraphics[clip,trim={0.5cm 0 0 0},scale=0.65]{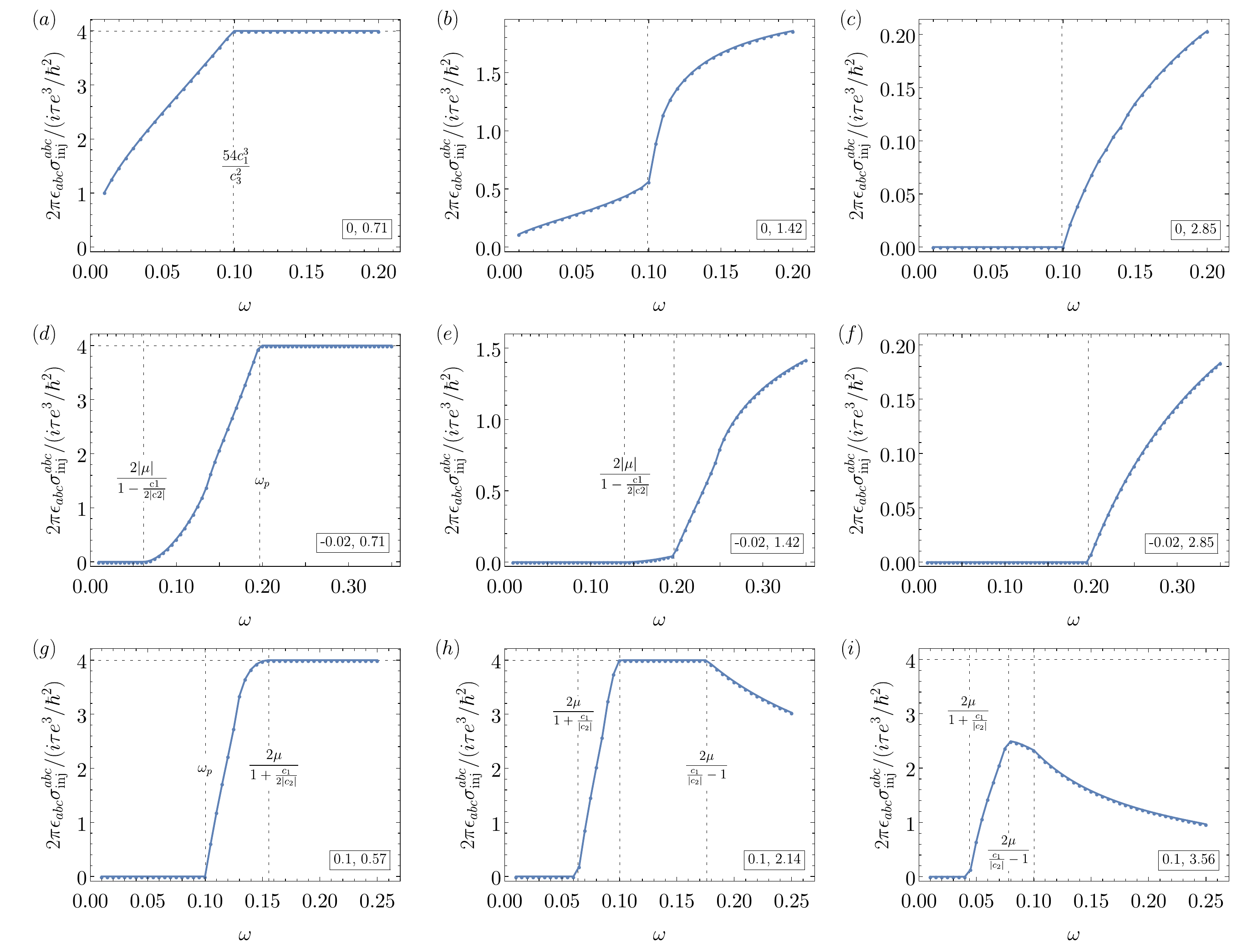}
\caption{Trace of the CPGE tensor for a single charge-4 Weyl point Eq.\eqref{eq:2bkpc4} with $c_1>0$, obtained from the numerical evaluation of integrals in Table~\ref{tab:c4} for different combinations of $\mu$ and $c_1/|c_2|$ (respective values shown in the inset). We have taken $c_1=0.0665,c_3=0.4$. (a)-(c) $\mu=0$, (d)-(f) $\mu<0$. In both these cases, perfect quantization is seen only when $c_1/|c_2|<1$. (g)-(i) $\mu>0$, perfect quantization is guaranteed for $c_1/|c_2|<1$ however, unlike previous two cases, it can also be seen for $c_1/|c_2|>1$ as long as $\frac{2\mu}{c_1/|c_2|-1} > \text{max}\left(\w_p,\frac{2\mu}{1+c_1/2|c_2|}\right)$, as shown in (h).} 
\label{figc4kp}
\end{figure*}

When $\mu=0$, quantization is seen only for $c_1/|c_2|<1$ starting at a frequency of $\frac{54c_1^3}{c_3^2}$ as shown in Fig.~\ref{figc4kp} (a). The situation for $1<c_1/|c_2|<2$ and $2<c_1/|c_2|$ is also shown in Fig.~\ref{figc4kp} (b) and (c), respectively. While the trace is non-zero for any finite frequency in the former case, it turns non-zero only after $\frac{54c_1^3}{c_3^2}$ for the latter.

For $\mu<0$, a perfect quantization is again only possible for $c_1/|c_2|<1$. When this is the case, the trace becomes non-zero after $\text{min}\left(\w_p,\frac{2|\mu|}{1-\frac{c_1}{2|c_2|}}\right)$ and reaches $\pm 4$ at $\text{max}\left(\w_p,\frac{2|\mu|}{1-\frac{c_1}{|c_2|}}\right)$ where $\w_p$ is the unique real positive root of $(\w-2|\mu|)^3-54\frac{c_1^3}{c_3^2}\w^2 = 0$. For $1<c_1/|c_2|<2$, the trace becomes non-zero after $\text{min}\left(\w_p,\frac{2|\mu|}{1-\frac{c_1}{2|c_2|}}\right)$ while for $2<c_1/|c_2|$, this happens after $\w_p$. The three cases are shown in Fig.~\ref{figc4kp} (d), (e) and (f), respectively. 
%

\begin{figure*}[t!]
\includegraphics[scale=0.58]{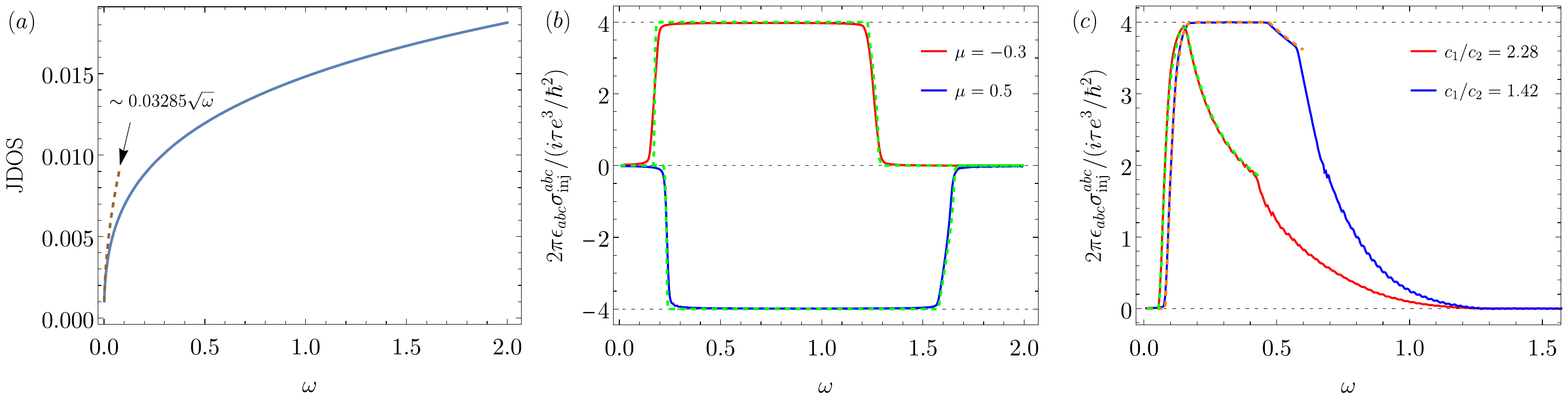}
\caption{(a) JDOS for a single charge-4 Weyl point with $\mu=0, c_1=0.0665, c_2=0.4668, c_3=4$ ($c_1/c_2<1$). (b) The red and blue curves capture the CPGE quantization for the two-band tight binding model Eq.\eqref{eq:2btbc4} with $\mu=-0.3,0.5$, respectively. Note that $\mu_\Gamma=\mu+6c_1, \mu_R=\mu-6c_1$. The corresponding dashed green curve is obtained by evaluating expressions from Table \ref{tab:c4} for each node separately and then adding the results. We have used the same $c_1, c_2, c_3$ from before. (c) Results with $c_1/c_2>1$, $\mu=-0.3$ (same $c_1,c_3$ as before). The dashed curves show contribution from the node closer to $E=0$ (obtained using results from Table~\ref{tab:c4}) for each case and are discontinued after contribution from the other node becomes non-zero. }
\label{figc4tb}
\end{figure*}

When $\mu>0$, we are presented with a wider range of possibilities for observing quantization. We find that, irrespective of the $c_1/|c_2|$ value, the trace becomes non-zero after $\text{min}\left(\w_p,\frac{2\mu}{1+\frac{c_1}{|c_2|}}\right)$ where $\w_p$ is now the unique real positive root of the cubic equation $(\w-2\mu)^3+54\frac{c_1^3}{c_3^2}\w^2 = 0$. For $c_1/|c_2|<1$, it goes on to reach a saturation value of $\pm 4$ at $\text{max}\left(\w_p,\frac{2\mu}{1+\frac{c_1}{2|c_2|}}\right)$, as shown in Fig.~\ref{figc4kp}(g). Interestingly for $\mu>0$, quantization becomes possible even for $1<c_1/|c_2|$ provided $\text{max}\left(\w_p,\frac{2\mu}{1+\frac{c_1}{2|c_2|}}\right)<\frac{2\mu}{\frac{c_1}{|c_2|}-1}$. When this condition is met, perfect quantization is seen but only for a finite window of energy $\w$ satisfying $\text{max}\left(\w_p,\frac{2\mu}{1+\frac{c_1}{2|c_2|}}\right)<\w<\frac{2\mu}{\frac{c_1}{|c_2|}-1}$ which is shown in Fig.~\ref{figc4kp}(h). The situation when no quantization is possible for $\mu>0$ is shown in Fig.~\ref{figc4kp}(i). It is clear that while $|c_1/c_2|$ plays a crucial role, $c_1,c_2,c_3$ and $\mu$ intricately determine the behavior of the CPGE trace and its quantization. Note that the plots in Fig.~\ref{figc4kp} have been obtained by numerically evaluating the integrals in Table~\ref{tab:c4}. We have included the JDOS plot at zero doping in Fig.~\ref{figc4tb}(a). As shown in the figure, the JDOS has a $\sqrt{\w}$ behavior going towards zero frequency. Note that the JDOS result from Table~\ref{tab:cn} also predicts a $\sqrt{\w}$ dependence if we take $n=4$. This seems more like a coincidence as that model still has a linearly dispersing band along $k_z$, very different from the C-4 model in Eq.\eqref{eq:2bkpc4}.

We would like to point out that so far the results presented assume $c_1>0$. It turns out that we can continue to use the same results for a charge-4 node with $c_1<0$ (and a chemical potential $\mu$) by treating it as a $|c_1|$ node with chemical potential $-\mu$. With this small but important extension, our analysis covers all the possible cases.

For completeness, we also compute the CPGE trace using the full tight-binding model Eq.\eqref{eq:2btbc4}, and the results are shown in Fig.~\ref{figc4tb}(b),(c). Note that in this model, when going from $\Gamma$ to $R$, we find $(c_1,c_2,c_3)\rightarrow (-c_1,-c_2,-c_3)$. Since $c_1$ turns negative, when using results from Table~\ref{tab:c4}, we treat the $(c_1<0,\mu_R)$ $R$-node as a $(|c_1|,-\mu_R)$ node. Also, since $c_3$ flips sign too, $\chi_\Gamma=-\chi_R$, as expected.

In Fig.~\ref{figc4tb}(b), we chose $c_1/|c_2|<1$ and therefore expect both nodes to show perfect quantization. Since $\mu\neq 0$, the two nodes will show quantization starting at different frequencies which results in an overall finite quantization window. The dashed curves represent the sum of contributions from the two nodes based on the low energy result from Table~\ref{tab:c4} (as remarked earlier, this makes sense here because at no point in the energy range under consideration do the contributions from both nodes become non-constant simultaneously).

In Fig.~\ref{figc4tb}(c), we chose $1<c_1/|c_2|$ and $\mu=-0.3$. This gives $\mu_\Gamma=0.1$ and $\mu_R=-0.7$. Since the node closer to zero energy falls under the $\mu>0$ category when using the low-energy results, we can choose $c_1,c_2,c_3$ such that it shows quantization for a finite window (blue curve) or no quantization at all (red curve). For the former, we also ensure that the contribution from other node starts only after the end of the quantization window. The dashed curves show contribution from the $\Gamma$-node in both cases. In both Fig.~\ref{figc4tb}(b) and (c), we find excellent agreement between the tight binding and low energy results, showing that the higher order terms are not at play in this parameter range and can be neglected.

%
\section{Conclusion and discussion}
In summary, we have presented a comprehensive and unified study of the second-order dc response in tilted multi-Weyl systems with a focus on the roles played by tilt ($W$) and doping ($\mu$). For charges $n$=1, 2 and 3, we have derived analytical expressions for shift and injection conductivity using a low energy continuum model and then compared its predictions against more realistic two- and four-band tight binding models of time-reversal broken systems for the charge-2 case. 

Beyond the extremely important CPGE quantization, we also report other features of the photogalvanic response arising mainly from the finite tilt and band curvatures. We systematically investigated the role of tilt, band curvatures, and higher bands in deciding the shift and injection current conductivities of multi-Weyl semimetals. We find that  in TRS broken multi-Weyl semimetals, finite tilt can lead to non-zero injection current from linearly polarized light which not only provides a probe for the tilt direction but can possibly also provide a way to engineer the injection current by using strain or some other mechanism which controls the tilt of Weyl nodes.

We have also provided the first complete analysis of the photogalvanic response in charge-4 WSM based on a low-energy two-band model, covering all  possibilities arising from different combinations of model parameters and the chemical potential. Although C-4 WSMs do not have a tilt in the usual sense (like the other three charges which have linear dispersing band in at least one direction), the ratio $|c_1/c_2|$ plays a similar role, and together with $c_1^3/c_3^2$ and $\mu$ determines nature of the response. Within the confines of the low-energy model, our results help point out exactly when CPGE quantization can be seen in C-4 WSMs.

We believe that the new approach we have taken here to study C-4 would find applications in studying many other optical responses as well. For example, it can easily be extended to the study of SHG and first-order conductivity for the low energy two band model. In principle it should work with any quantity that requires evaluating a $k$-space integral with $f_{21}\delta(\w_{21}-\w)$ term in it at $T=0$ K.

\section{Acknowledgement}
S.C. would like to acknowledge the funding 
from the National Science Foundation through the
Center for Dynamics and Control of Materials: an
NSF MRSEC under Cooperative Agreement No. DMR-
1720595. G.A.F. acknowledges additional support from
NSF DMR-2114825

\appendix
%
\section{Analytical expression for shift and injection conductivity tensors}
\label{appendix:cn}

We work with the low-energy effective Hamiltonian,
\begin{align}
 \HH_n &= \mqty(u_zk_z + u_tk_z - \mu    & \E0(\kt_x-i\zeta\kt_y)^n \\
              \E0(\kt_x+i\zeta\kt_y)^n   & -u_zk_z + u_tk_z - \mu),
\end{align} 
with eigenvalues
\begin{align}
 E_{n,\pm} &= u_tk_z - \mu\pm\E0\sqrt{(\kt_x^2+\kt_y^2)^n + u_z^2k_z^2/\E0^2}.
\end{align}

The domain for the integrals in Eq.\eqref{eq:shift_main}, Eq.\eqref{eq:inj_main} is determined by $f_{21}$ and $\delta (\w_{21}-\w) = \delta (2\E0\sqrt{(\kt_x^2+\kt_y^2)^n + u_z^2k_z^2/\E0^2}-\w)$. Let us focus on the delta function first. To simplify things, we split the integral in $k_x-k_y$ plane over the four quadrants: $\int\dd{k_x}\int\dd{k_y} = \int_{+}\dd{k_x}\int_{+}\dd{k_y} + \int_{+}\dd{k_x}\int_{-}\dd{k_y} + \int_{-}\dd{k_x}\int_{+}\dd{k_y} + \int_{-}\dd{k_x}\int_{-}\dd{k_y}$, and combine them into a single integral over the first quadrant by making substitutions $k_x = \pm k_0\sqrt{x}$ and $k_y = \pm k_0\sqrt{y}$ depending on the sign (both $x,y>0$). We also put $k_z=\frac{\E0}{u_z}z$. By making $x\rightarrow\frac{x-y}{\sqrt{2}}$ and $y\rightarrow\frac{x+y}{\sqrt{2}}$, we rotate the $x-y$ axis counterclockwise by $\pi/4$, and scale $z$ by $z\rightarrow 2^{n/4}z$. Finally, we let $x\rightarrow x^{2/n}$ to get $\delta( 2^{1+n/4}\E0\sqrt{x^2+z^2}-\w )$. These transformations also change the integration measure $\int_{\kv}\rightarrow \int\tfrac{2^{1/2+n/4} k_0^2\E0 x^{2/n-1} \dd{x}\dd{z}\dd{y}}{16\pi^3|u_z|n\sqrt{x^2-y^2}}$, with $x>0$ and $-x<y<x$ (we do the $y$ integral first with these limits). The delta function defines a circle in $xz$-plane which lets us use $x=r\cos{\theta}, z=r\sin{\theta}$ to obtain $\delta( r-\w/(2^{1+n/4}\E0))/(2^{1+n/4}\E0)$. 

Since we are taking the temperature to be zero, $f_{21}= \Theta(E_1)-\Theta(E_2)$ where, $\Theta$ is the Heaviside step function. Because of the condition put by the delta function, we have 
\begin{align}
 f_{21} &= \Theta\left(u_tk_z-\mu-\frac{\w}{2}\right) - \Theta\left(u_tk_z-\mu+\frac{\w}{2}\right).
\end{align}

Since we have assumed $\w>0$, the only non-zero value for $f_{21}$ is $-1$ when $\mu-\w/2<u_z k_z<\mu+\w/2$. Using coordinate transformations from before, this condition becomes 
\begin{gather}
 \frac{2\mu}{\w}-1<\frac{u_t}{u_z}\sin{\theta}<\frac{2\mu}{\w}+1, \\
 \frac{\text{sgn}\left(\frac{u_t}{u_z}\right)\frac{2\mu}{\w}-1}{W}<\sin{\theta}<\frac{\text{sgn}\left(\frac{u_t}{u_z}\right)\frac{2\mu}{\w}+1}{W}, 
\end{gather}
where $W=|u_t/u_z|$. The definitions for $\theta_1,\theta_2$ given in main text follow from this. With this, we can easily compute other ingredients of the integral from the eigenvalues and normalized eigenfunctions of $\HH_n$, and combine them to obtain analytical expressions for the JDOS, shift and injection conductivities.

%
\section{Higher order terms for charge-2 WSM}
\label{appendix:hoc}
Based on the higher order terms appearing in the expansion  of Eq.\eqref{eq:2btbc2} near its nodes, we look at the effect of including $(\tfrac{1}{2}(k_x^2+k_y^2) + u_m k_z^2)\s_z - \tfrac{1}{2}g u_z k_z^2 \s_0$ in Eq.\eqref{eq:2bkpc2} where $u_m=\tfrac{1}{2}M-1$. As before, we use a series of transformations to simplify the Dirac delta constraint. Key steps are as follows (with $\E0=1$), 
\begin{enumerate}
    \item $k_x \rightarrow\pm\sqrt{x}$, $k_y \rightarrow\pm \sqrt{y}$, $k_z\rightarrow\frac{u_z}{u_m}z$.
    \item $x\rightarrow\frac{1}{\sqrt{2}}(x-y)$, $y\rightarrow\frac{1}{\sqrt{2}}(x+y)$, $z\rightarrow \frac{1}{2}(\sqrt{z}-1)$.
    \item $z\rightarrow\frac{\sqrt{40}u_m}{u_z^2}z+1$, integrate out $y$ (from $-x,x$).
    \item $x\rightarrow\frac{x}{2\sqrt{5+\sqrt{5}}}-\frac{z}{2\sqrt{5-\sqrt{5}}}$, $z\rightarrow\frac{x}{2\sqrt{5+\sqrt{5}}}+\frac{z}{2\sqrt{5-\sqrt{5}}}$.
    \item $x\rightarrow\w\cos{\theta}$, $z\rightarrow\w\sin{\theta}$, integrate from $\theta_1,\theta_2$.
\end{enumerate}
Analytical expressions for JDOS, shift and injection conductivity tensors can be obtained as before. We still have $f_{21}=-1$, however, the condition that determines $\theta_1,\theta_2$ becomes, 
\begin{align}
\begin{split}
    2 \mu -\omega <\frac{u_t u_z}{u_m} \left(\sqrt{\frac{\w\sqrt{5} u_m\sin(\theta+\beta)}{u_z^2}+1}-1\right) \\
    -\frac{g u_z^3}{4 u_m^2} \left(\sqrt{\frac{\w\sqrt{5} u_m\sin(\theta+\beta)}{u_z^2}+1}-1\right)^2<2 \mu +\omega,
\end{split}
\end{align}
where $\beta=\arctan(\varphi-1)$, $-\frac{\pi}{2}-\arctan(\varphi)\leq\theta\leq\frac{\pi}{2}-\arctan(\varphi)$, and $\varphi$ is the golden ratio. Allowed values of $\theta$ can be found by solving this inequality numerically. When solutions turn out to be disjoint intervals, each interval defines its own $\theta_1,\theta_2$. The analytical expression is evaluated for each interval and then summed.

%
\section{Tilt and Zeeman terms for charge-2 WSM}
We can also include additional terms of the form $A(\kt_x^2+\kt_y^2)\s_0$ and $B\s_z$ into Eq.\eqref{eq:Hn}. These correspond to second-order tilt and Zeeman terms, respectively. The $B$ term only shifts the origin along $k_z$, modifying the  $k_z\rightarrow z$ transformation to $k_z=\frac{\E0 z - B}{u_z}$. It not difficult to see that these terms only affect $f_{21}$,
\begin{align}
\begin{split}
 f_{21} &= \Theta\left(\frac{A}{\E0}\frac{\w\cos{\theta}}{2}+\frac{u_t}{u_z}\frac{\w\sin{\theta}}{2}-\widetilde{\mu}-\frac{\w}{2}\right) \\
        &\qquad - \Theta\left(\frac{A}{\E0}\frac{\w\cos{\theta}}{2}+\frac{u_t}{u_z}\frac{\w\sin{\theta}}{2}-\widetilde{\mu}+\frac{\w}{2}\right), 
\end{split}
\end{align}
where $\widetilde{\mu}=\mu+Bu_t/u_z$. Since $\w>0$, we have $f_{21}=-1$ when,
\begin{gather}
 \frac{2\widetilde{\mu}}{\w}-1<\frac{A}{\E0}\cos{\theta}+\frac{u_t}{u_z}\sin{\theta}<\frac{2\widetilde{\mu}}{\w}+1.
\end{gather}
By defining $\widetilde{W}=\sqrt{A^2/\E0^2+u_t^2/u_z^2}$, $\sin{\phi}=\frac{A/\E0}{\widetilde{W}}$, and $\cos{\phi}=\frac{|u_t/u_z|}{\widetilde{W}}$, we obtain
\begin{gather}
\begin{split}
 \frac{\text{sgn}\left(\tfrac{u_t}{u_z}\right)\frac{2\widetilde{\mu}}{\w}-1}{\widetilde{W}}<\sin(\theta+\text{sgn}\left(\tfrac{u_t}{u_z}\right)\phi) < & \\
 & \hspace{-1.5cm} \frac{\text{sgn}\left(\tfrac{u_t}{u_z}\right)\frac{2\widetilde{\mu}}{\w}+1}{\widetilde{W}},
\end{split} 
\end{gather}
with $-\pi/2\leq\theta,\phi\leq\pi/2$. Also, we define $\alpha=\text{sgn}\left(\tfrac{u_t}{u_z}\right)\phi$ and $\widetilde{\varphi}_p=\frac{1}{\widetilde{W}}\left(\text{sgn}\left(\frac{u_t}{u_z}\right)\frac{2\widetilde{\mu}}{\w}+(-1)^p\right)$ with $p=1,2$. The inequality becomes $\widetilde{\varphi}_1<\sin(\theta+\alpha)<\widetilde{\varphi}_2$ which can be solved for the minimum ($\widetilde{\theta_1}$) and maximum ($\widetilde{\theta_2}$) allowed values of $\theta$. These can be obtained as the left and right end points of the intervals,
\begin{align}
 (\widetilde{\theta_1},\widetilde{\theta_2}) &= \begin{dcases}
                       (\pi/2,\pi/2) ,\quad 1<\tphi_1, 1<\tphi_2  \\
                       (-\pi/2,-\pi/2) ,\quad \tphi_1<-1, \tphi_2<-1  \\
                       I_0 ,\quad \tphi_1<-1, 1<\tphi_2  \\
                       I_0\cap I_2 ,\quad 0<\tphi_1<1, 1<\tphi_2  \\
                       I_0\cap I_1 ,\quad -1<\tphi_1<0, 1<\tphi_2  \\
                       I_0\cap I_3 ,\quad \tphi_1<-1, -1<\tphi_2<0  \\
                       I_0\cap I_4 ,\quad \tphi_1<-1, 0<\tphi_2<1  \\
                       I_0\cap I_1\cap I_4 ,\quad -1<\tphi_1<0<\tphi_2<1  \\
                       I_0\cap I_1\cap I_3 ,\quad -1<\tphi_1<\tphi_2<0  \\
                       I_0\cap I_2\cap I_4 ,\quad 0<\tphi_1<\tphi_2<1  \\
                      \end{dcases}
\end{align}
where,
\begin{align*}
 I_0 &= (-\pi/2,\pi/2), \\
 I_1 &= (-\pi-\A,-\pi-\A-\arcsin{\tphi_1})\cup (-\A+\arcsin{\tphi_1},\pi-\A), \\
 I_2 &= (-\A+\arcsin{\tphi_1},\pi-\A-\arcsin{\tphi_1}), \\
 I_3 &= (-\pi-\A-\arcsin{\tphi_2},-\A+\arcsin{\tphi_2}), \\
 I_4 &= (-\pi-\A,-\A+\arcsin{\tphi_2}) \cup (\pi-\A-\arcsin{\tphi_2},\pi-\A).
\end{align*}

%
\section{Sign changing of $\s^{xyz}_\mathrm{shift}$ for $W>2$}
\label{appendix:signchange}

For $W<2$, we found that the sign change occurred at $\w=2|\mu|$ when one angle was $\pm\pi/2$ and the other zero. However, when $W>2$, $(\theta_2,\theta_1)$ cannot take either $(0,-\pi/2)$, or $(\pi/2,0)$. Finding the point of sign change now requires us to seek other solutions of $\sin{\theta_2}\cos^2{\theta_2} - \sin{\theta_1}\cos^2{\theta_1}=0$. Converting cosine into sine we get $(\sin{\theta_2}-\sin{\theta_1})(\sin^2{\theta_2} + \sin{\theta_2}\sin{\theta_1} + \sin^2{\theta_1}-1)=0$. Let's look for solutions other than $\theta_2=\theta_1$, $(0,-\pi/2)$, and $(\pi/2,0)$. We can solve for $(\sin{\theta_2},\sin{\theta_1})$ to get,
\begin{align}
 (\sin{\theta_2},\sin{\theta_1}) &= \begin{dcases}
                     \left(\tfrac{-x - \sqrt{4-3x^2}}{2},x\right),\, -1<x<\tfrac{-1}{\sqrt{3}} \\
                     \left(x,\tfrac{-x - \sqrt{4-3x^2}}{2}\right),\, \tfrac{-1}{\sqrt{3}}<x<0 \\
                     \left(\tfrac{-x + \sqrt{4-3x^2}}{2},x\right),\, 0<x<\tfrac{1}{\sqrt{3}} \\
                     \left(x,\tfrac{-x + \sqrt{4-3x^2}}{2}\right),\, \tfrac{1}{\sqrt{3}}<x<1        
                    \end{dcases}
\end{align}
Using definitions of $\theta_1,\theta_2$, we solve for $\w$ by eliminating $x$ to obtain $\w=2|\mu|\sqrt{\frac{3}{W^2-1}}$. Note that for $W=2$, this gives $\w=2|\mu|$ as expected.

%
\section{Analytical results for charge-4 WSM}
\label{appendix:c4}
The delta function constraint $\delta(\w-\w_{21})$ translates to
\begin{align}
\begin{split}
 &4|c_2|\bigg(k_x^4+k_y^4+k_z^4 - k_x^2k_y^2-k_y^2k_z^2-k_z^2k_x^2 \\
 &\quad\hspace{3cm}+ \left(\frac{c_3}{2c_2}\right)^2 k_x^2k_y^2k_z^2 \bigg)^\frac{1}{2}=\w.     
\end{split} 
\end{align}
To simplify this, we use the following transformations,
\begin{enumerate}
    \item $k_x \rightarrow\pm\sqrt{x}$, $k_y \rightarrow\pm \sqrt{y}$, $k_z\rightarrow\pm\sqrt{z}$ (reduce the integral to $x,y,z>0$ octant).
    \item $x\rightarrow\frac{1}{\sqrt{2}}(x-y)$, $y\rightarrow\frac{1}{\sqrt{2}}(x+y)$.
    \item Integrate out $y$ (from $-x,x$). To do this, we need to find roots of the equation $\w=\left(2 c_3^2 z \left(x^2-y^2\right)+8 c_2^2 \left(\left(x-\sqrt{2} z\right)^2+3 y^2\right)\right)^{\frac{1}{2}}$. The condition for existence of real roots satisfying $-x\leq y\leq x$ is given by (post step 4 substitution).
    \begin{align}
    \begin{split}
     &\left(4 c_2^2 \left(x^2-x z+z^2\right)-c_1^2\right) \big(-c_3^2 x^2 \w z \\ 
     &\hspace{2.5cm} -8 c_2^2 c_1 (x-2 z)^2+8 c_1^3\big)>0.
    \end{split}\label{eq:c4_newcondition}
    \end{align}
    \item $x\rightarrow\frac{\w}{2\sqrt{2}c_1}x$, $z\rightarrow\frac{\w}{2c_1}z$ .
\end{enumerate}
Using these transformations along with the eigenvalues and normalized eigenfunctions of Eq.\eqref{eq:2bkpc4}, we simply Eq.\eqref{eq:shift_main}, Eq.\eqref{eq:inj_main}, Eq.\eqref{eq:jdos} to obtain the expressions shown in Table \ref{tab:c4} (the shift conductivities are zero).

\begin{figure}[t!]
\includegraphics[scale=0.67]{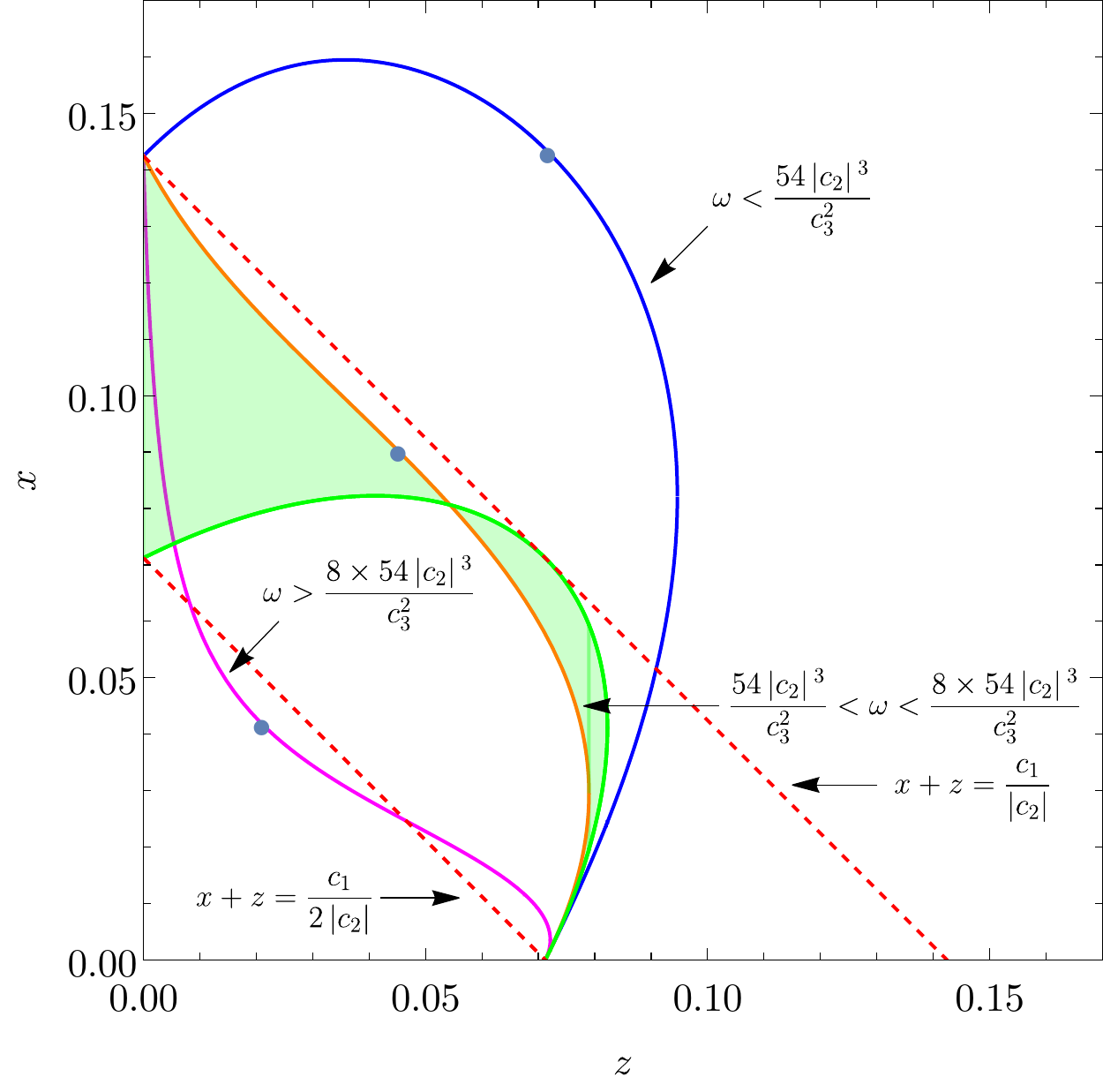}
\caption{The green area is the region defined by Eq.\eqref{eq:c4_newcondition} (we have used $c_1=0.0665, c_2=0.4668, c_3=4$, $\w=0.4$, but the features shown here are quite general). The blue, orange and magenta curves correspond to $-c_3^2 x^2 \w z -8 c_2^2 c_1 (x-2 z)^2+8 c_1^3=0$ for $\w=0.1, 0.4$ and $4.0$, respectively. The green curve represents the ellipse $4 c_2^2 \left(x^2-x z+z^2\right)-c_1^2=0$. The blue dots are placed at $\left(\frac{2^{1/3}c_1}{|c_3|^{2/3}\w^{1/3}},\frac{2\times 2^{1/3}c_1}{|c_3|^{2/3}\w^{1/3}}\right)$.}
\label{figE2}
\end{figure}

Behavior of $\frac{2\pi}{i\tau e^3/\hbar^2} \epsilon_{abc}\s^{abc}$ is determined by the interplay between conditions set by $\Theta\left(-x-z+1+\frac{2\mu}{\w}\right)$, $\Theta\left(x+z+1-\frac{2\mu}{\w}\right)$ and Eq.\eqref{eq:c4_newcondition}. Since $c_1>0$ by choice and $c_2,c_3$ appear only as their squares, the analysis of region defined by Eq.\eqref{eq:c4_newcondition} becomes quite general. To understand this, let us focus on the curves $4 c_2^2 \left(x^2-x z+z^2\right)-c_1^2=0$ and $-c_3^2 x^2 \omega  z-8 c_2^2 c_1 (x-2 z)^2+8 c_1^3=0$ for $x,z>0$. They intersect the $x$-axis at $x=c_1/2|c_2|$ and $x=c_1/|c_2|$, respectively, but cross the $z$-axis together at $z=c_1/2|c_2|$ (intercepts are independent of $\w$). Tangents to these curves with slope $-1$ are important. For the ellipse this happens at $(z=c_1/2|c_2|,x=c_1/2|c_2|)$, the tangent has equation $x+z=c_1/|c_2|$. For the second curve we have several cases. For $\w<48\frac{|c_2|^3}{c_3^2}$, there is only one such tangent at ($z=\frac{2^{1/3}c_1}{|c_3|^{2/3}\w^{1/3}},x=\frac{2\times 2^{1/3}c_1}{|c_3|^{2/3}\w^{1/3}}$), described by $x+z=\frac{54^{1/3}c_1}{|c_3|^{2/3}\w^{1/3}}$. For larger $\w$, there is another tangent with slope $-1$, but its presence is of no consequence to our analysis. The important thing to note is that $x+z=\frac{54^{1/3}c_1}{|c_3|^{2/3}\w^{1/3}}$ is completely sandwiched between $x+z=c_1/|c_2|$ and $x+z=c_1/2|c_2|$ for $54\frac{|c_2|^3}{c_3^2}<\w<8\times 54\frac{|c_2|^3}{c_3^2}$. These features are illustrated in Fig.~\ref{figE2}. With these key observations in mind, we now analyze the $\mu=0$, $\mu<0$, and $\mu>0$ cases separately.

For $\mu=0$, the theta function constraints reduce to $x+z<1$. When $c_1/|c_2|<1$, the CPGE trace is non-zero for any finite $\w$ and becomes $\pm 4$ after $\frac{54c_1^3}{c_3^2}$. When $c_1/|c_2|>1$, some portion of Eq.\eqref{eq:c4_newcondition} is always left out and we do not get perfect quantization. For $1<\frac{c_1}{|c_2|}<2$, the trace is non-zero for any finite $\w$ where as for $\frac{c_1}{|c_2|}>2$, this happens only after $\frac{54c_1^3}{c_3^2}$.

For $\mu<0$, the condition set by $\Theta\left(x+z+1-\frac{2\mu}{\w}\right)$ is always satisfied where as $\Theta\left(-x-z+1+\frac{2\mu}{\w}\right)$ requires $x+z<1-\frac{2|\mu|}{\w}$. An important thing to note here is that the term $1-\frac{2|\mu|}{\w}\in (-\infty,1)$. We are only interested when it lies in $(0,1)$ which happens for $\w>2|\mu|$. Since it can ever only reach 1, full overlap with region Eq.\eqref{eq:c4_newcondition} is possible if $c_1/|c_2|<1$, the condition to get perfect quantization. When this condition is met, the amount of overlap between $x+z<1-\frac{2|\mu|}{\w}$ and Eq.\eqref{eq:c4_newcondition} is determined by solutions to equations $1-\frac{2|\mu|}{\w}=\frac{c_1}{|c_2|}$, $1-\frac{2|\mu|}{\w}=\frac{c_1}{2|c_2|}$, and $1-\frac{2|\mu|}{\w}=\frac{54^{1/3}c_1}{|c_3|^{2/3}\w^{1/3}}$. The last equation can be rewritten as $(\w-2|\mu|)^3-54\frac{c_1^3}{c_3^2}\w^2=0$. This cubic equation never has three real roots. Since the product of its roots is $8|\mu|^3>0$, the only real root, $\w_p$, is always positive. Note that $\w_p>2|\mu|$. The CPGE trace becomes non-zero after $\text{min}\left(\w_p,\frac{2|\mu|}{1-\frac{c_1}{2|c_2|}}\right)$, and saturates to $\pm 4$ after $\text{max}\left(\w_p,\frac{2|\mu|}{1-\frac{c_1}{|c_2|}}\right)$. When $1<c_1/|c_2|<2$, the trace is non-zero after $\text{min}\left(\w_p,\frac{2|\mu|}{1-\frac{c_1}{2|c_2|}}\right)$ where as for $c_1/|c_2|>2$, this happens after $\w_p$ (it never reaches $\pm 4$ in either case).

For $\mu>0$, the possibilities become even more interesting. $\Theta\left(-x-z+1+\frac{2\mu}{\w}\right)\Theta\left(x+z+1-\frac{2\mu}{\w}\right)$ sets bounds on the integration region, requiring $\frac{2\mu}{\w}-1<x+z<\frac{2\mu}{\w}+1$. The term $1+\frac{2\mu}{\w}\in (1,\infty)$, which means that if $c_1/|c_2|>1$, a portion of Eq.\eqref{eq:c4_newcondition} will necessarily be left out for $\w>\frac{2\mu}{\frac{c_1}{|c_2|}-1}$ (perfect quantization still possible for smaller energies). Now, the solutions to equations $\frac{2\mu}{\w}-1=\frac{c_1}{|c_2|}$, $\frac{2\mu}{\w}-1=\frac{c_1}{2|c_2|}$, and $\frac{2\mu}{\w}-1=\frac{54^{1/3}c_1}{|c_3|^{2/3}\w^{1/3}}$ become crucial in determining the amount of region Eq.\eqref{eq:c4_newcondition} available for integration. The last equation can be rewritten as the cubic equation $(\w-2\mu)^3+54\frac{c_1^3}{c_3^2}\w^2=0$. The product of its roots is $8\mu^3>0$ which means that when two roots are complex (conjugate pair), the real root must be positive. However, when all roots are real, there are two possibilities $-$ one positive two negative roots or three positive roots. It turns out the latter case is not possible because the condition for all roots being real is $\mu<4\frac{c_1^3}{c_3^2}$ where as for all roots to be positive, $\mu>9\frac{c_1^3}{c_3^2}$. Thus, we always get exactly one positive root, $\w_p$. Note that $\w_p<2\mu$ in this case. The CPGE trace becomes non-zero after $\text{min}\left(\w_p,\frac{2\mu}{1+\frac{c_1}{|c_2|}}\right)$. For $c_1/|c_2|<1$, it goes on to reach a saturation value of $\pm 4$ after $\text{max}\left(\w_p,\frac{2\mu}{1+\frac{c_1}{2|c_2|}}\right)$. When $c_1/|c_2|>1$, we see quantization for $\text{max}\left(\w_p,\frac{2\mu}{1+\frac{c_1}{2|c_2|}}\right)<\w<\frac{2\mu}{\frac{c_1}{|c_2|}-1}$. Perfect quantization is not possible when $\frac{2\mu}{\frac{c_1}{|c_2|}-1}<\text{max}\left(\w_p,\frac{2\mu}{1+\frac{c_1}{2|c_2|}}\right)$.

\bibliography{mWSM.bib}
\end{document}